\documentclass[english,aps,reprint,superscriptaddress]{revtex4-1}
\usepackage[T1]{fontenc}
\usepackage[latin9]{inputenc}
\usepackage{verbatim}
\usepackage{amsmath}
\usepackage{graphicx}

\makeatletter
\usepackage{babel}

\makeatother

\usepackage{babel}
\begin{document}
\title{Mass splitting and spin alignment for $\phi$ mesons in a magnetic
field in NJL model}
\author{Xin-Li Sheng}
\email{xls@mail.ccnu.edu.cn}

\affiliation{Key Laboratory of Quark and Lepton Physics (MOE) and Institute of
Particle Physics, Central China Normal University, Wuhan, 430079,
China}
\author{Shu-Yun Yang}
\affiliation{Key Laboratory of Quark and Lepton Physics (MOE) and Institute of
Particle Physics, Central China Normal University, Wuhan, 430079,
China}
\author{Yao-Lin Zou}
\affiliation{Key Laboratory of Quark and Lepton Physics (MOE) and Institute of
Particle Physics, Central China Normal University, Wuhan, 430079,
China}
\author{Defu Hou}
\email{houdf@mail.ccnu.edu.cn}

\affiliation{Key Laboratory of Quark and Lepton Physics (MOE) and Institute of
Particle Physics, Central China Normal University, Wuhan, 430079,
China}
\begin{abstract}
Based on the Nambu-Jona-Lasinio (NJL) model, we develop a framework
for calculating the spin alignment of vector mesons and applied it
to study $\phi$ mesons in a magnetic field. We calculate mass spectra
for $\phi$ mesons and observe mass splitting between the longitudinally
polarized state and transversely polarized states. The $\phi$
meson in a thermal equilibrium system is preferred to occupy the state
with spin $\lambda=0$ than those with spin $\lambda=\pm1$, because
the former state has a smaller energy. As a consequence, we conclude
that the spin alignment will be larger than 1/3 if one measures along
the direction of the magnetic field, which is qualitatively consistent
with the recent STAR data. Around the critical temperature $T_{C}=150$
MeV, the positive deviation from 1/3 is proportional to the square
of the magnetic field strength, which agrees with the result from
the non-relativistic coalescence model. Including the anomalous magnetic
moments for quarks will modify the dynamical masses of quarks and
thus affect the mass spectra and spin alignment of $\phi$ mesons.
The discussion of spin alignment in the NJL model may help us better
understand the formation of hadron's spin structure during the chiral
phase transition. 
\end{abstract}
\maketitle

\section{Introduction }

Non-central relativistic heavy-ion collisions provide a unique opportunity
to study quantum chromodynamics (QCD) matter in a strong magnetic
field \citep{Rafelski:1975rf}. The hot and dense matter created in
collisions is known as the quark-gluon plasma (QGP), which evolves
with time and cools down to the hadronic phase at the freeze-out time.
In Au-Au collisions at the Relativistic Heavy Ion Collider (RHIC)
or Pb-Pb collisions at the Large Hadron Collider (LHC), the magnetic
field perpendicular to the reaction plane can reach $5\,m_{\pi}^{2}\sim10^{18}$
Gauss ($m_{\pi}$ is the pion mass) or even larger \citep{Kharzeev:2007jp,Skokov:2009qp,Kharzeev:2013jha,Deng:2012pc,Tuchin:2013apa}.
Such a strong magnetic field is mainly generated by spectators in
the colliding nuclei. It drops fastly with time, but the existence
of medium electrical conductivity will extend its lifetime \citep{McLerran:2013hla,Gursoy:2014aka,Tuchin:2014iua,Li:2016tel,Chen:2021nxs,Yan:2021zjc,Wang:2021oqq}.
Therefore it may have sizeable contributions to many phenomena, for
example, the chiral magnetic effect \citep{Kharzeev:2007jp,Fukushima:2008xe,Son:2009tf},
the $\Lambda$'s polarization \citep{Liang:2004ph,Becattini:2016gvu,STAR:2017ckg},
and the charge-odd directed flow \citep{Gursoy:2014aka,Das:2016cwd,Gursoy:2018yai,Dubla:2020bdz,Zhang:2022lje}.
On the other hand, the electromagnetic fields also have event-by-event
fluctuations, which are still significantly large even in the late
stage of collisions \citep{Bzdak:2011yy,Deng:2012pc,Siddique:2021smf}.
The fluctuating fields thus contribute to phenomena at the freeze-out
time, such as the magnetic catalysis \citep{Klimenko:1991he,Gusynin:1994re,Gusynin:1995nb,Miransky:2002rp,Endrodi:2019zrl},
the inverse magnetic catalysis \citep{Bali:2011qj,Endrodi:2019zrl},
and the phase structure of the QGP \citep{Gusynin:1994re,Fraga:2008um,Mizher:2010zb,Andersen:2014xxa}.

Recently, the STAR collaboration has measured the $\phi$ and $K^{*0}$
meson's spin alignment along the out-of-plane direction and observes
a significant positive deviation from 1/3 for the $\phi$ meson \citep{STAR:2022fan}.
The spin alignment refers to the 00-element of the normalized spin
density matrix for a vector meson with spin-1 \citep{Liang:2004xn,Tang:2018qtu}.
The positive derivation from 1/3 observed in experiments indicates
that the spin of $\phi$ meson is preferred to align in the reaction
plane. According to the quark coalescence model, the spin alignment
of vector meson is induced by polarizations of its constituent quarks
\citep{Liang:2004xn,Yang:2017sdk} and thus have various sources such
as the vorticity field \citep{Liang:2004xn,Yang:2017sdk,Xia:2020tyd},
the electromagnetic field \citep{Yang:2017sdk}, the helicity polarization
\citep{Gao:2021rom}, the turbulent color field \citep{Muller:2021hpe},
the shear stress \citep{Li:2022vmb,Wagner:2022gza}, and the strong
force field \citep{Sheng:2019kmk,Sheng:2020ghv,Sheng:2022ffb,Sheng:2022wsy}.
Among these works, only the fluctuations of the strong force field
successfully reproduce the experiment data \citep{Sheng:2022wsy}.
Since the strong force field in \citep{Sheng:2022wsy} has the same
structure as the classical electromagnetic field, one naturally expects
that fluctuations of electromagnetic fields, rather than their event-average
values, also contribute to the spin alignment of vector mesons.

In this work, we study the spin alignment of $\phi$ meson in a constant
magnetic field using the three flavor Nambu-Jona-Lasinio (NJL) model
\citep{Nambu:1961fr,Nambu:1961tp,Klimt:1989pm,Vogl:1989ea,Klevansky:1992qe,Buballa:2003qv,Volkov:2005kw,Fukushima:2008wg}.
Such a field configuration can be straightforwardly extended to the
case of a space-time dependent magnetic field with the typical length
of its inhomogeneity much larger than the typical hadron size. In
the NJL model, gluons are integrated out and quarks interact via local
four-fermion interactions, which have the form that keeps the chiral
symmetry. Mesons are treated as quantum fluctuations beyond a constant
mean-field and their propagators are introduced through the random
phase approximation by the resummation of quark bubbles \citep{Klevansky:1992qe,Hatsuda:1994pi,Buballa:2003qv}.
The mass spectra for mesons are given by the poles, 
of their propagators. Within the framework of
magnetized NJL model, the spectra of light-flavor mesons, including
$\sigma$, $\pi^{0}$, $\pi^{\pm}$, $\omega$, $\rho^{0}$, and $\rho^{\pm}$,
have attracted a lot of interest \citep{Liu:2014uwa,Avancini:2015ady,Avancini:2016fgq,Mao:2018dqe,Coppola:2018vkw,Chaudhuri:2019lbw,Xu:2020yag,Wei:2020xfd,Yang:2021hud},
but few works focus on the $\phi$ meson. One can refer to \citep{Andersen:2014xxa,Miransky:2015ava,Cao:2021rwx}
for recent reviews on the NJL model in a strong magnetic field. In
this manuscript, we observe the splitting between masses of $\phi$
mesons in different spin states, which is induced by the magnetization
of the constituent quark and antiquark. In a hot and thermal equilibrium
system, the mass splitting leads to different spin-dependent equilibrium
distributions and thus corresponds to a nontrivial spin alignment.
We also study the effect of quark anomalous magnetic moments (AMM)
considering that constituent quarks have different magnetic moments
compared with free quarks \citep{Brekke:1987cc,Chang:2010hb,Fayazbakhsh:2014mca,Ayala:2015bgv,Chaudhuri:2019lbw,Xu:2020yag}.
The AMMs are included in the fermion Hamiltonian by putting a new
term $q_{f}\kappa_{f}F_{\mu\nu}\sigma^{\mu\nu}/2$, where $F^{\mu\nu}$
is the electromagnetic field tensor, $\sigma^{\mu\nu}\equiv(i/2)[\gamma^{\mu},\gamma^{\nu}]$,
and $q_{f}$, $\kappa_{f}$ are the charge and the AMM for a quark
with flavor $f=u,d,s$. The AMMs change the dynamical masses of quarks
and therefore affect the spectra and spin alignment of the $\phi$
meson.

This manuscript is organized as follows. In Sec. \ref{sec:Nambu-Jona-Lasinio-model}
we review the theoretical framework for the three flavor NJL model
and numerically calculate quark dynamical masses. Then in Sec. \ref{sec:vector meson}
we give analytical formulas for the vector meson's propagator, the
spectral function, and the spin alignment. Numerical results for
$\phi$ mesons are given in Sec. \ref{sec:Numerical-results}. We
then repeated the calculations in the presence of nonzero AMMs in
Sec. \ref{sec:Effect-of-anomalous}. Finally, in Sec. \ref{sec:Summary}
we summarize our findings and conclude. %

\section{Nambu-Jona-Lasinio model for quarks\label{sec:Nambu-Jona-Lasinio-model}}

\subsection{Theoretical framework}

In order to describe a strongly-interaction quark matter, we use the
three-flavor Nambu-Jona-Lasinio model with scalar and vector channels
of four-fermion interactions \citep{Klevansky:1992qe,Hatsuda:1994pi,Buballa:2003qv,Volkov:2005kw},
\begin{align}
\mathcal{L}_{\text{eff}} & =\mathcal{L}_{q}+G_{S}\sum_{a=0}^{8}\left[(\overline{\psi}\lambda_{a}\psi)^{2}+(\overline{\psi}i\gamma_{5}\lambda_{a}\psi)^{2}\right]\nonumber \\
 & -G_{V}\sum_{a=0}^{8}\left[(\overline{\psi}\gamma_{\mu}\lambda_{a}\psi)^{2}+(\overline{\psi}i\gamma_{\mu}\gamma_{5}\lambda_{a}\psi)^{2}\right]\nonumber \\
 & -K\left\{ \text{det}_{f}\left[\overline{\psi}(1+\gamma_{5})\psi\right]+\text{det}_{f}\left[\overline{\psi}\left(1-\gamma_{5}\right)\psi\right]\right\} \,,\label{eq:effective Lagrangian}
\end{align}
where $\psi=(\psi_{u},\psi_{d},\psi_{s})$ are Dirac spinors for $u$,
$d$, and $s$ quarks, respectively, $\lambda_{a}$ with $a=1,2,\cdots,8$
are Gell-Mann matrices, and $\lambda_{0}=\sqrt{2/3}I_{0}$ with $I_{0}$
being the identity matrix in the color space. The last term in Eq. (\ref{eq:effective Lagrangian}) is the
six-quark Kobayashi-Maskawa - 't Hooft interaction that breaks the
$U_{A}(1)$ symmetry \citep{tHooft:1976rip}. Here $G_{S}$ and $G_{V}$
are coupling constants for scalar and vector interactions, respectively.
The Lagrangian $\mathcal{L}_{q}$ for quarks in an external electromagnetic
field is given by 
\begin{equation}
\mathcal{L}_{q}=\sum_{f=u,d,s}\overline{\psi}_{f}\left(i\gamma_{\mu}D_{f}^{\mu}-m_{f}\right)\psi_{f}\,,
\end{equation}
where $m_{f}$ denotes current mass for quarks with flavor $f=u,d,s$.
The covariant derivative is $D_{f}^{\mu}\equiv\partial^{\mu}+iq_{f}A^{\mu}$
with $q_{f}$ being the quark charges and $A^{\mu}$ being the gauge
potential for the external electromagnetic field. Under the mean-field
approximation, the Lagrangian becomes 
\begin{eqnarray}
\mathcal{L}_{\text{MF}} & = & \sum_{f=u,d,s}\overline{\psi}_{f}\left(i\gamma_{\mu}D_{f}^{\mu}-M_{f}\right)\psi_{f}\nonumber \\
 &  & -2G_{S}\sum_{f=u,d,s}\sigma_{f}^{2}+4K\sigma_{u}\sigma_{d}\sigma_{s}\,,\label{eq:MF_Lagrangian}
\end{eqnarray}
where $\sigma_{f}$ is the quark chiral condensate $\sigma_{f}\equiv\left\langle \overline{\psi}_{f}\psi_{f}\right\rangle $.
Here we only consider the chiral condensate and set all other possible
condensates to zeros. The dynamical mass $M_{f}$ is related
to $\sigma_{f}$ as 
\begin{equation}
M_{f}\equiv m_{f}-4G_{S}\sigma_{f}+2K\prod_{f^{\prime}\neq f}\sigma_{f^{\prime}}\,,\label{eq:Dynamical masses}
\end{equation}
where the last term arises from the 't Hooft interaction.

We consider quarks in a constant magnetic field. Without loss of generality,
we assume the magnetic field is along the positive $z$-direction
and take the Landau gauge $A^{\mu}=(0,0,Bx,0)$. For each flavor of
quark, it is straightforward to derive the Dirac equation from the
Lagrangian (\ref{eq:MF_Lagrangian}). The Dirac equation can be analytically
solved by applying the Ritus method \citep{Ritus:1972ky,Ritus:1978cj},
resulting in the following dispersion relation for the $n$-th Landau
level, 
\begin{equation}
E=\pm E_{f,n}(p_{z})=\pm\sqrt{p_{z}^{2}+M_{f}^{2}+2n|q_{f}B|}\,,\label{eq:eigenenergies}
\end{equation}
where the momentum perpendicular to the $z$-direction is quantized
as the Landau levels, while the longitudinal momentum is not restricted.
Here $B>0$ denotes the magnetic field strength and $q_{f}$ is the
electric charge of a quark with flavor $f$, with $q_{u}=(2/3)e$,
$q_{d}=q_{s}=(-1/3)e$, and $e$ being the elementary charge. The
positive and negative energies are related to particles and antiparticles,
respectively. Using Eq. (\ref{eq:eigenenergies}), the quark grand
thermodynamic potential $\Omega_{f}$ can be written as 
\begin{align}
\Omega_{f} & =\frac{N_{c}|q_{f}B|}{4\pi^{2}}\sum_{n,s}\int dp_{z}\Biggl\{\frac{E_{f,n}(p_{z})}{2}\nonumber \\
 & +2T\ln\left[1+e^{-E_{f,n}(p_{z})/T}\right]\Biggr\}\,,\label{eq:quark_potential}
\end{align}
where $T$ is the temperature and $N_{c}=3$ is the degeneracy of
color. The summation in Eq. (\ref{eq:quark_potential}) runs over
$s=+$ for the lowest Landau level $n=0$ and $s=\pm$ for other Landau
levels $n=1,2,3,\cdots$. The total grand potential for the whole
system includes $\Omega_{f}$ and the mean field part, which is given
by 
\begin{equation}
\Omega=\sum_{f=u,d,s}\left(2G_{S}\sigma_{f}^{2}-\Omega_{f}\right)+4K\sigma_{u}\sigma_{d}\sigma_{s}\,.
\end{equation}
The quark condensates $\sigma_{f}$ and the corresponding quark masses
$M_{f}$ are then calculated by minimizing the grand potential, $\partial\Omega/\partial\sigma_{f}=0$.

\subsection{Numerical results \label{subsec:Numerical-quark-mass}}

Since the NJL model is non-renormalizable, it is necessary to include
a regularization scheme for the divergent momentum integrals in Eqs.
(\ref{eq:quark_potential}). Since a sharp three-momentum cutoff will
lead to nonphysical oscillations in the presence of a magnetic field,
we choose a Pauli-Villas regularization scheme \citep{Pauli:1949zm}.
Any function of $M_{f}$ is replaced by a summation, 
\begin{equation}
f(M_{f})\rightarrow f_{\text{P.V.}}(M_{f})=\sum_{j=0}^{3}c_{j}f\left(\sqrt{M_{f}^{2}+j\Lambda^{2}}\right)\,,
\end{equation}
with $c_{0}=1$, $c_{1}=-3$, $c_{2}=3$, and $c_{3}=-1$. We take
the parameter set given in \citep{Carignano:2019ivp}, 
\begin{eqnarray}
m_{u,d}=10.3\,\text{MeV}, & m_{s}=236.9\,\text{MeV}, & \Lambda=0.7812\,\text{GeV},\nonumber \\
G_{S}\Lambda^{2}=4.90, & K\Lambda^{5}=129.8,
\end{eqnarray}
which are obtained by fitting vacuum values for the pion decay constant
and masses of pion, kaon, $\eta^{\prime}$, while fixing the vacuum
mass for light-quarks to $M_{u,d}=325$ MeV. For the strange quark,
this set of parameters leads to a dynamical mass $M_{s}=554$ MeV in
the vacuum in absence of the magnetic field. 

\begin{figure}
\includegraphics[width=8cm]{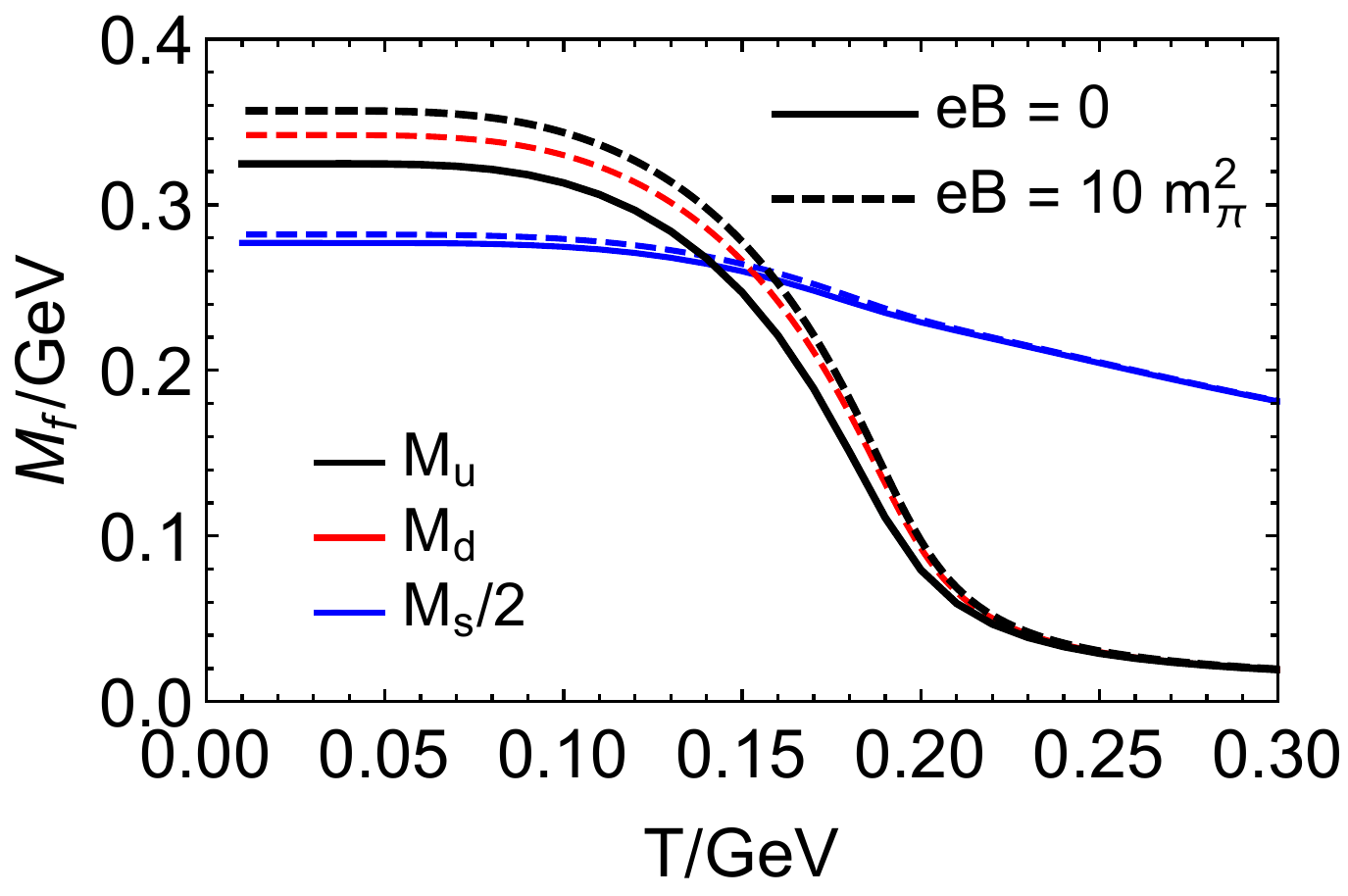}

\caption{\label{fig:Dynamical-masses}Dynamical masses as functions of the
temperature for $u$ quark (black lines), $d$ quark (red lines),
and $s$ quark (blue lines). The magnetic field strength is set
to $eB=0$ (solid lines) or $eB=10\,m_{\pi}^{2}$ (dashed lines).}
\end{figure}

We first focus on dynamical masses for quarks, which are related
to the chiral condensates $\sigma_{f}$ as given in Eq. (\ref{eq:Dynamical masses}).
The condensates $\sigma_{f}$ are order parameters for the chiral
phase transition. In the chiral symmetry breaking phase, $\sigma_{f}\neq0$
and thus quarks have nonvanishing dynamical masses. In the chiral
symmetry restored phase, $\sigma_{f}=0$ and quark dynamical masses
reduce to their current masses. The behavior of dynamical masses for
$u$, $d$, and $s$ quarks as functions of the temperature is shown
in Fig. \ref{fig:Dynamical-masses}. Here we choose two sets of values
for the magnetic field strength: solid lines for $eB=0$ and dashed
lines for $eB=10\,m_{\pi}^{2}$. We observe that the chiral phase
transition is a cross-over with critical temperature around $T_{C}\approx150$
MeV. Compared to the case with $eB=0$, a nonzero magnetic field, $eB=10\,m_{\pi}^{2}$,
corresponds to larger quark masses and slightly higher $T_{C}$, which
is the behavior of the magnetic catalysis. We also observe that the
$u$ and $d$ quarks have identical masses when $eB=0$, but have
different masses in a nonzero magnetic field. That is because the
difference in their electric charges, $q_{u}=(2/3)e$ and $q_{d}=(-1/3)e$,
leads to different magnetic energies and breaks the symmetry between
light-flavor quarks.

\begin{figure}
\includegraphics[width=8cm]{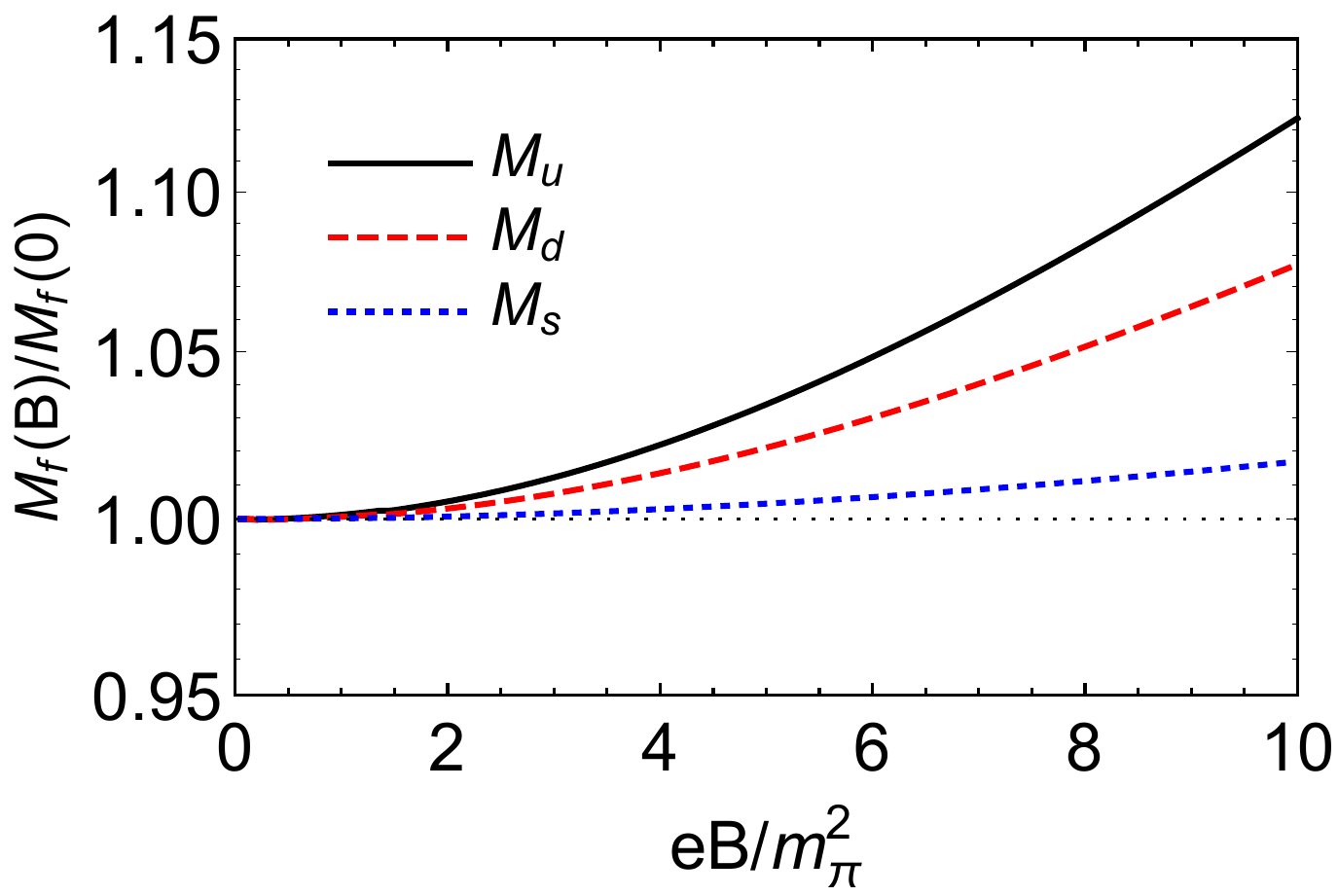}

\caption{\label{fig:Dynamical-mass-eB-1}Dynamical masses as functions of the
magnetic field strength at $T=150$ MeV for $u$, $d$, and $s$
quarks, denoted by the black solid line, the red dashed line, and the blue
dotted line, respectively. The masses are normalized by the corresponding values in absence of magnetic field.}
\end{figure}

In order to explicitly show the magentic field dependence, we plot
in Fig. \ref{fig:Dynamical-mass-eB-1} the ratio of quark masses as
functions of the field strength to those in absence of the magnetic
field. Here we fix the temperature at the ordinary critical temperature
$T=150$ MeV. We find that quark masses grow with an increasing magnetic
field, which is the phenomena of magnetic catalysis. The $u$ quark
mass is more affected by the magnetic field than the $d$ quark mass
since $|q_{u}|>|q_{d}|$. On the other hand, the $s$ quark is less
affected by the magnetic field because $s$ quark has a larger dynamical
mass than $u,\,d$ quarks. 

\section{Spin alignment for $\phi$ mesons in NJL model \label{sec:vector meson}}

\subsection{Propagator for $\phi$ mesons}

\begin{figure}
\includegraphics[width=7cm]{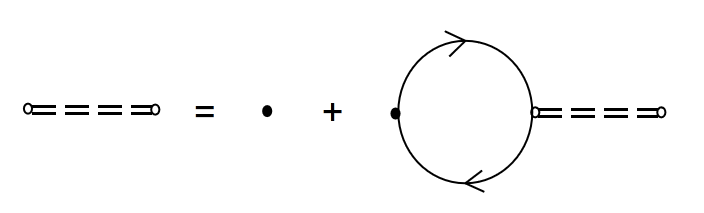}

\caption{\label{fig:Dyson-equation}Dyson-Schwinger equation for meson's propagator.
Solid lines denote quark propagators, dashed lines denote meson propagator,
 black dots denotes four-quark vertexes, and white dots denotes the quark-meson vertexes.}
\end{figure}

In the NJL model, mesons are described as excitations beyond the mean
field. The progator of a meson is obtained by taking the quark bubble
summation in the random phase approximation, corresponding to the
Dyson-Schwinger equation shown in Fig. \ref{fig:Dyson-equation}.
For the vector meson channel, the Dyson-Schwinger equation is given
by 
\begin{equation}
D^{\mu\nu}(k)=4G_{V}\Delta^{\mu\nu}(k)+4G_{V}\Delta^{\mu\alpha}(k)\Sigma_{\alpha\beta}(k)D^{\beta\nu}(k)\,,\label{eq:Dyson equation}
\end{equation}
where $D^{\mu\nu}$ is the vector meson's propagator and $\Sigma^{\mu\nu}$
is the self-energy tensor. The projection operator
$\Delta^{\mu\nu}(k)\equiv g^{\mu\nu}-k^{\mu}k^{\nu}/k^{2}$ ensures
the Ward identity $k_{\mu}D^{\mu\nu}(k)=0$. Using the quark propator
$S_{f}(p)$, the self-energy at one-quark-loop level is given by 
\begin{equation}
\Sigma^{\mu\nu}(k)=-iN_{c}\int\frac{d^{4}p}{(2\pi)^{4}}\text{Tr}\left[\gamma^{\mu}S_{f}(k+p)\gamma^{\nu}S_{f^{\prime}}(p)\right]\,,\label{eq:self-energy}
\end{equation}
where the quark flavors $f$ and $f^{\prime}$ depend on the type
of considered meson. In this work we focus the $\phi$ meson and thus
we take $f=f^{\prime}=s$ in Eq. (\ref{eq:self-energy}). Following
Ref. \citep{Miransky:2015ava}, the quark propagator in momentum space
is expressed as follows,
\begin{equation}
S_{f}(p)=ie^{-p_{\perp}^{2}/|q_{f}B|}\sum_{n=0}^{\infty}\frac{(-1)^{n}D_{n}^{f}(p)}{(k^{0})^{2}-[E_{f,n}(k_{z})]^{2}+i\epsilon}\label{eq:fermion propagator}
\end{equation}
where the poles are shifted above or below the real axis by an infinite
small imaginary part $\pm i\epsilon$ such that $S_{f}(p)$ denotes
the Feynmann propagator. The residue at each pole energy is determined
by the function in the numerator,
\begin{eqnarray}
D_{n}^{f}(p) & = & 2(p^{0}\gamma^{0}-p_{z}\gamma^{3}+M_{f})\left[\mathcal{P}_{+}L_{n}\left(\frac{2p_{\perp}^{2}}{|q_{f}B|}\right)\right.\nonumber \\
 &  & \left.-\mathcal{P}_{-}L_{n-1}\left(\frac{2p_{\perp}^{2}}{|q_{f}B|}\right)\right]\nonumber \\
 &  & +4{\bf p}_{\perp}\cdot\boldsymbol{\gamma}_{\perp}L_{n-1}^{1}\left(\frac{2p_{\perp}^{2}}{|q_{f}B|}\right)\label{eq:Df-n}
\end{eqnarray}
where the projection operators are $\mathcal{P}_{\pm}=[1\pm i\gamma^{1}\gamma^{2}\text{sgn}(q_{f}B)]/2$
and $L_{n}^{i}(x)$ are associated Laguerre polynomials with $L_{n}(x)\equiv L_{n}^{0}(x)$. 

As spin-1 particles, the vector mesons have three spin states, $\lambda=0,\,\pm1$.
Taking the spin quantization direction as the $z$-direction in the
meson's rest frame, which is the same as the direction of magnetic
field, we have the following spin polarization vectors, 
\begin{eqnarray}
\boldsymbol{\epsilon}_{0} & = & (0,0,1)\,,\nonumber \\
\boldsymbol{\epsilon}_{+1} & = & -\frac{1}{\sqrt{2}}(1,i,0)\,,\nonumber \\
\boldsymbol{\epsilon}_{-1} & = & \frac{1}{\sqrt{2}}(1,-i,0)\,,\label{eq:spin polarization vectors}
\end{eqnarray}
corresponding to $\lambda=0,\pm1$, respectively. By taking a Lorentz
boost, we derive the covariant form of spin polarization vectors,
\begin{equation} \label{eq:pol-vectors covariant}
\epsilon^{\mu}(\lambda,k)=\left(\frac{{\bf k}\cdot\boldsymbol{\epsilon}_{\lambda}}{m_{V}},\boldsymbol{\epsilon}_{\lambda}+\frac{{\bf k}\cdot\boldsymbol{\epsilon}_{\lambda}}{m_{V}(\omega+m_{V})}{\bf k}\right)\,,
\end{equation}
where $k^{\mu}\equiv(\omega,{\bf k})$ is the four-momentum and $m_{V}=\sqrt{\omega^{2}-{\bf k}^{2}}$
is the mass of the vector meson. It is easy to check that these vectors
are perpendicular to $k^{\mu}$, $k_{\mu}\epsilon^{\mu}(\lambda,k)=0$,
and they are properly normalized as $\epsilon^{\mu\ast}(\lambda,k)\epsilon_{\mu}(\lambda^{\prime},k)=-\delta_{\lambda\lambda^{\prime}}$.
They also form a complete basis as $\sum_{\lambda=0,\pm1}\epsilon^{\mu\ast}(\lambda,k)\epsilon^{\nu}(\lambda,k)=-\Delta^{\mu\nu}(k)$.
Then the meson propagator $D^{\mu\nu}(k)$ can be cast into the following
form, 
\begin{equation}
D^{\mu\nu}(k)=\sum_{\lambda,\lambda^{\prime}=0,\pm1}\epsilon^{\ast\mu}(\lambda,k)\epsilon^{\nu}(\lambda^{\prime},k)D_{\lambda\lambda^{\prime}}(k)\,,
\end{equation}
where the element $D_{\lambda\lambda^{\prime}}(k)$ is a Lorentz invariant
function that is derived by projecting $D^{\mu\nu}(k)$ onto $\epsilon_{\mu}(\lambda,k)\epsilon_{\nu}^{\ast}(\lambda^{\prime},k)$,
\begin{equation}
D_{\lambda\lambda^{\prime}}(k)=\epsilon_{\mu}(\lambda,k)\epsilon_{\nu}^{\ast}(\lambda^{\prime},k)D^{\mu\nu}(k)\,.
\end{equation}
Similarly, projecting the Dyson-Schwinger equation in (\ref{eq:Dyson equation})
onto $\epsilon_{\mu}(\lambda,k)\epsilon_{\nu}^{\ast}(\lambda^{\prime},k)$
gives the equation for $D_{\lambda\lambda^{\prime}}(k)$,%
\begin{equation}
D_{\lambda\lambda^{\prime}}(k)=-4G_{V}\delta_{\lambda\lambda^{\prime}}-4G_{V}\sum_{\lambda_{1}=0,\pm1}\Sigma_{\lambda\lambda_{1}}(k)D_{\lambda_{1}\lambda^{\prime}}(k)\,,\label{eq:D-S equation spin space}
\end{equation}
where the self-energy matrix element in the spin space, $\Sigma_{\lambda\lambda^{\prime}}(k)$,
is defined as
\begin{equation}
\Sigma_{\lambda\lambda^{\prime}}(k)=\epsilon_{\mu}(\lambda,k)\epsilon_{\nu}^{\ast}(\lambda^{\prime},k)\Sigma^{\mu\nu}(k)\,.
\end{equation}
Equation (\ref{eq:D-S equation spin space}) has the following formal
solution, 
\begin{equation}
D_{\lambda\lambda^{\prime}}(k)=-\left[\frac{4G_{V}}{1+4G_{V}\Sigma(k)}\right]_{\lambda\lambda^{\prime}}\,.\label{eq:reduced propagator}
\end{equation}
where $1$ and $\Sigma(k)$ in the denominator are short-handed notations
for the $3\times3$ unit matrix and the matrix $\Sigma_{\lambda\lambda^{\prime}}(k)$. 

\subsection{Spectral function and spin alignment}

In a thermal equilibrium system, the meson's propagator can be expressed
in the spectral representation as \citep{Kapusta:2006pm} 
\begin{equation} \label{eq:spectral representation}
D^{\mu\nu}(k)=n_{\text{BE}}(\omega/T)\frac{1}{\pi}\text{Im}D^{\mu\nu}(k)\,,
\end{equation}
where $n_{\text{BE}}(\omega/T)=1/(e^{\omega/T}-1)$ is the Bose-Einstein
distribution. The density matrix for the vector meson is the given
by 
\begin{equation}
\overline{\rho}_{\lambda\lambda^{\prime}}({\bf k})=\int d\omega\,2\omega\,n_{\text{BE}}(\omega/T)\xi_{\lambda\lambda^{\prime}}(k)\,,\label{eq:density matrix}
\end{equation}
where the spectral function is derived from the imaginary part of
the full propagator, 
\begin{equation}
\xi_{\lambda\lambda^{\prime}}(k)\equiv\frac{1}{\pi}\epsilon_{\mu}(\lambda,k)\epsilon_{\nu}^{\ast}(\lambda^{\prime},k)\text{Im}D^{\mu\nu}(k)\,.\label{eq:spectra function}
\end{equation}
In Eq. (\ref{eq:density matrix}), we integrate over $\omega$ so
that the diagonal element $\overline{\rho}_{\lambda\lambda}({\bf k})$
has definite physical meaning of particle number for vector mesons with spin
$\lambda$ and three-momentum ${\bf k}$. The spin alignment is then
given by the $00$-element of the normalized density matrix, 
\begin{equation}
\rho_{00}({\bf k})\equiv\frac{\overline{\rho}_{00}({\bf k})}{\sum_{\lambda=0,\pm1}\overline{\rho}_{\lambda\lambda}({\bf k})}\,.\label{eq:spin alignment}
\end{equation}
The result in a constant magnetic field can be evaluated by using
Eqs. (\ref{eq:self-energy}), (\ref{eq:fermion propagator}), and (\ref{eq:spectral representation}) - (\ref{eq:spin alignment}).
When calculating the self-energy in (\ref{eq:self-energy}), we substitute
the energy integral with a summation over Matsubara frequencies at
finite temperature \citep{Kapusta:2006pm}. This allows us to study
the meson properties at finite temperature.

We emphasize that Eqs. (\ref{eq:spectral representation}) - (\ref{eq:spin alignment})
are universal formulas, which can be applied in calculating momentum-dependent
spin alignments along any measuring direction. However, it will significantly
simplify our calculation to focus on a static meson ${\bf k}={\bf 0}$
and choose the measuring direction as the $z$-direction. The corresponding
spin polarization vectors as given in Eqs. (\ref{eq:spin polarization vectors}) and (\ref{eq:pol-vectors covariant}).
Due to the rotational invariance in the $x-y$ plane, one can also
prove that $\Sigma_{\lambda\lambda^{\prime}}$, $D_{\lambda\lambda^{\prime}}$,
and the density matrix $\overline{\rho}_{\lambda\lambda^{\prime}}({\bf 0})$
are diagonal in the spin space, 
\begin{equation}
\overline{\rho}_{\lambda\lambda^{\prime}}({\bf 0})=\left(\begin{array}{ccc}
\overline{\rho}_{11} & 0 & 0\\
0 & \overline{\rho}_{00} & 0\\
0 & 0 & \overline{\rho}_{-1,-1}
\end{array}\right)\,,
\end{equation}
where the states with $\lambda=\pm1$ are degenerate, $\overline{\rho}_{-1,-1}=\overline{\rho}_{11}$.
In general, if the measuring direction is characterized by Euler angles
$(\alpha,\,\beta,\,\gamma)$, the density matrix is calculated by
performing a rotation in spin space, 
\begin{align}
 & \overline{\rho}_{\lambda\lambda^{\prime}}({\bf 0};\alpha,\beta,\gamma)\nonumber \\
 & \ \ =\sum_{\lambda_{1},\lambda_{2}}R_{\lambda\lambda_{1}}(\alpha,\beta,\gamma)\overline{\rho}_{\lambda_{1}\lambda_{2}}({\bf 0})R_{\lambda_{2}\lambda^{\prime}}^{-1}(\alpha,\beta,\gamma)\,,
\end{align}
where $R_{\lambda\lambda^{\prime}}(\alpha,\beta,\gamma)$ is the spin-1
representation of the rotation with Euler angles $(\alpha,\,\beta,\,\gamma)$.
Here $\overline{\rho}_{\lambda_{1}\lambda_{2}}({\bf 0})$ is the density
matrix when measuring along the $z$-direction. A straightforward
calculation shows that the spin alignment is independent to Euler
angles $\alpha$ and $\gamma$, 
\begin{equation}
\rho_{00}({\bf 0};\alpha,\beta,\gamma)=\frac{\overline{\rho}_{00}({\bf 0})\cos^{2}\beta+\overline{\rho}_{11}({\bf 0})\sin^{2}\beta}{\overline{\rho}_{00}({\bf 0})+2\overline{\rho}_{11}({\bf 0})}\,.
\end{equation}
Defining the spin alignment in the magnetic field direction as 
\begin{equation}
\rho_{00}^{B}({\bf 0})\equiv\frac{\rho_{00}({\bf 0})}{\rho_{00}({\bf 0})+2\rho_{11}({\bf 0})}\,,
\end{equation}
we derive that 
\begin{equation}
\rho_{00}({\bf 0};\alpha,\beta,\gamma)=\frac{1}{2}\left\{ 1-\rho_{00}^{B}({\bf k})+\left[3\rho_{00}^{B}({\bf k})-1\right]\cos^{2}\beta\right\} \,,\label{eq:general result}
\end{equation}
which only depends on $\rho_{00}^{B}$ and the angle between the direction
of the magnetic field and the measuring direction. For a fluctuating
magnetic field, one has to take an average over the $\beta$-angle
and the field strength. If the field does not have a preferred direction,
one can prove that the average $\left\langle \rho_{00}({\bf 0};\alpha,\beta,\gamma)\right\rangle =1/3$,
as expected. If the fluctuations are anisotropic in space, the average
spin alignment will deviate from 1/3, as predicted in Refs. \citep{Sheng:2019kmk,Sheng:2020ghv,Sheng:2022ffb,Sheng:2022wsy}.

\section{Numerical results for $\phi$ mesons \label{sec:Numerical-results}}

The property of vector meson depends on the coupling strength $G_{V}$
for the vector channel in the Lagrangian (\ref{eq:effective Lagrangian}).
In our calculation, we take 
\begin{equation}
G_{V}\Lambda^{2}=-4.67\,,
\end{equation}
which is determined by fitting the $\phi$ meson's vacuum mass $M_{\phi}=1.02$
GeV in the absence of magnetic field.

\subsection{Mass spectra for $\phi$ meson \label{subsec:Spectra functions}}

Using quark masses as functions of the temperature and the magnetic
field strength discussed in Sec. \ref{subsec:Numerical-quark-mass},
we are then able to calculate the vector meson's spectral functions
from Eqs. (\ref{eq:self-energy}), (\ref{eq:fermion propagator}),
and (\ref{eq:spectra function}). In this work, we focus on the vector
$\phi$ meson, which has constituent quarks $s$ and $\bar{s}$. At
the one-loop level, the self-energy for the $\phi$ meson depends
on the propagator of $s$ quark, but does not depend on propagators
of $u$ and $d$ quarks. We set the spin quantization direction parallel
to the magnetic field and therefore the spectral function is diagonal
$\xi_{\lambda\lambda^{\prime}}=\text{diag(\ensuremath{\xi_{+1},\,\xi_{0}},\,\ensuremath{\xi_{-1}})}$,
and spin states $\lambda=\pm1$ are degenerate, $\xi_{+1}=\xi_{-1}$.
In this work, we only focus on static $\phi$ mesons, i.e., the three-momenta
of meson are set to zeros, ${\bf k={\bf 0}}$.

\begin{figure}
\includegraphics[width=8cm]{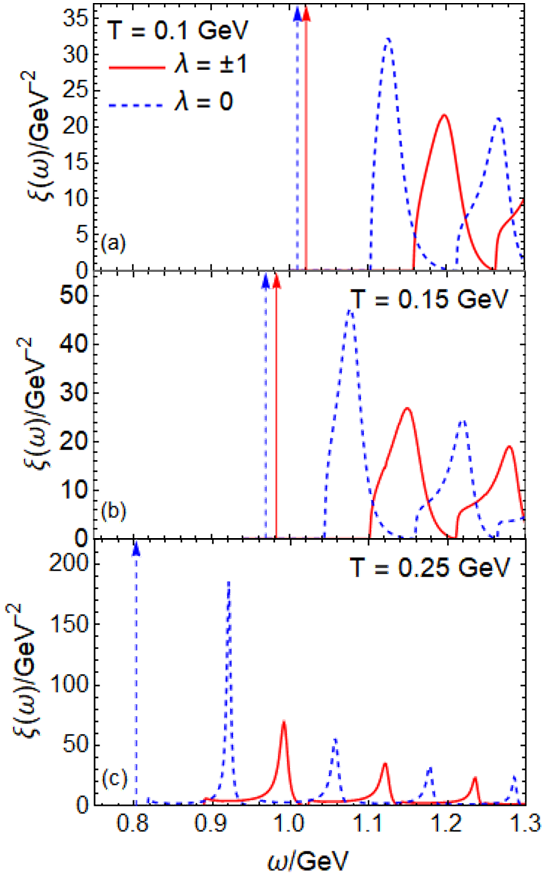}

\caption{\label{fig:Mass-spectra}Spectral functions for $\phi$ mesons in
a constant magnetic field with $eB=5m_{\pi}^{2}$, at $T=100$ MeV
[panel (a)], $T=150$ MeV [panel (b)], and
$T=250$ MeV [panel (c)]. Here red solid lines denote
spectral functions for $\phi$ mesons with spin state $\lambda=\pm1$
and blue dashed lines denote the spectral function of $\lambda=0$.
Red and blue arrows correspond to delta functions.}
\end{figure}

In order to clearly show the influence of the magnetic field to the
spectral function, we choose the magnetic field strength $eB=5\,m_{\pi}^{2}$
and plot the mass spectra for the $\phi$ meson at temperatures $T=100$
MeV, $150$ MeV, and $250$ MeV in Fig. \ref{fig:Mass-spectra}. In
general, the spectral function can be separated into a delta-function
part and a continuum part as 
\begin{equation}
\xi_{\lambda}(\omega,{\bf 0})=\delta(\omega^{2}-M_{\phi,\lambda}^{2})+\xi_{\lambda}^{\ast}(\omega)\,.\label{eq:spectra-separation}
\end{equation}
The delta-function part is identified as a stable bound state that
corresponds to a real on-shell $\phi$ meson with mass $M_{\phi,\lambda}$.
The continuum part $\xi_{\lambda}^{\ast}(\omega)$ is related to unstable
resonance excitations. In Fig. \ref{fig:Mass-spectra}, the delta-functions
are plotted as arrows. We observe a mass splitting between bound states
with $\lambda=0$ and $\lambda=\pm1$ in Fig. \ref{fig:Mass-spectra}
(a) and (b), indicating that the longitudinally polarized $\phi$
meson and the transversely polarized $\phi$ meson have different
energies. The bound state masses and the thresholds for the continuum
drop with increasing $T$, which is the result of the decreasing $s$
quark mass. The bound states will dissociate when $T$ is large enough,
which is the Mott transition \citep{Mott:1968nwb,Hufner:1996pq,Costa:2002gk,Blaschke:2016sqn,Mao:2019avr}.
In this work, the $\phi$ meson self-energy is purelly contributed
by the $s$ quark loop since we only consider the coupling between
the $\phi$ meson and the $s$ quark. In a thermal medium, the physical
$\phi$ meson was observed to have a large broadening \citep{Ishikawa:2004id,KEK-PS-E325:2005wbm,CLAS:2009kjz},
which may arise from $\phi N$ interations and is beyond the scope
of our discussion. We also observe in Fig. \ref{fig:Mass-spectra}
that the continuum parts contains several well-separated peaks. One
can understand the multi-peak structure of the spectral function in
Fig. \ref{fig:Mass-spectra} as follows. In the presence of a magnetic
field, the dispersion relation of $s$ quark is quantized as Landau
levels given in Eq. (\ref{eq:eigenenergies}). Supposing the constituent
$s$ and $\bar{s}$ quarks inside a $\phi$ meson are at Landau levels
$n_{1}$ and $n_{2}$, respectively, the angular momentum conservation
demands that the $\phi$ meson with $\lambda=0$ must be consist of
quark and antiquark with $n_{1}=n_{2}$, and the $\phi$ meson with
$\lambda=\pm1$ correspond to $n_{1}=n_{2}\pm1$. Different sets of
$n_{1}$ and $n_{2}$ give different resonance peaks as shown in Fig.
\ref{fig:Mass-spectra}.

\begin{figure}
\includegraphics[width=8cm]{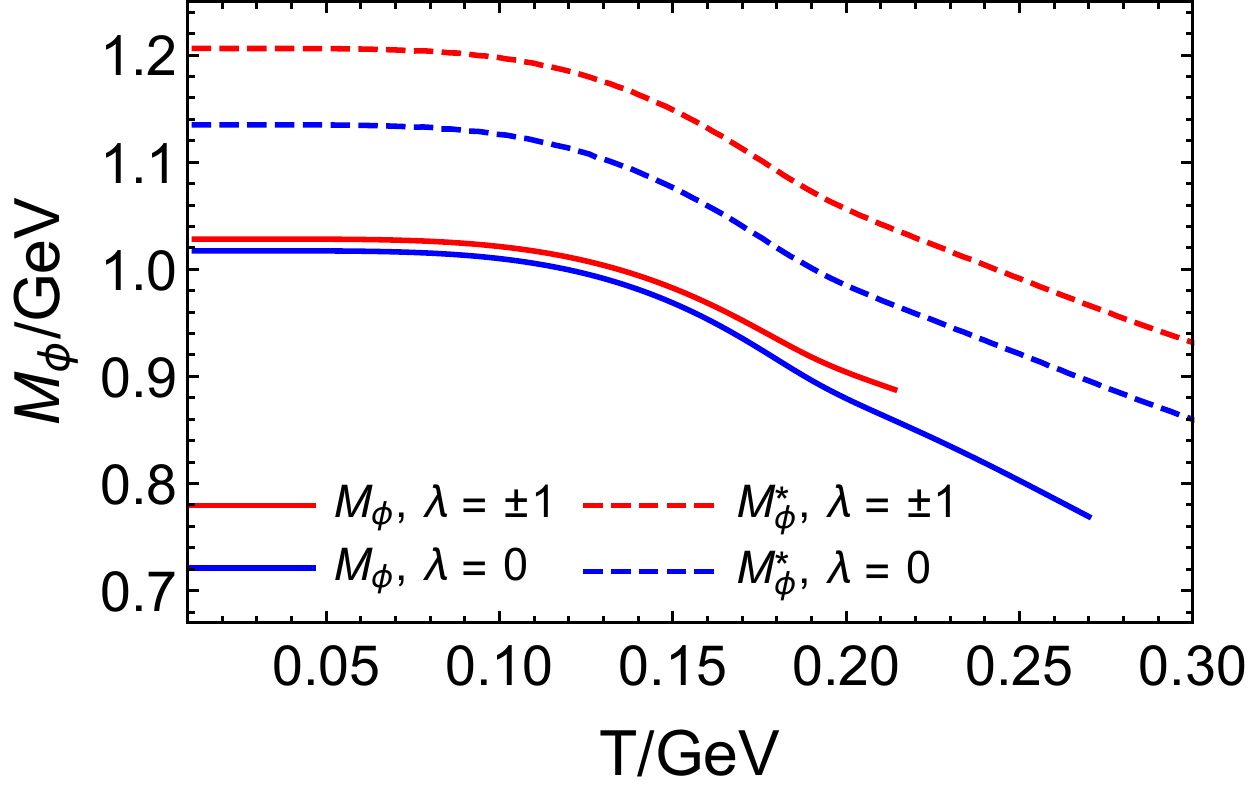}

\caption{\label{fig:Vector-Mass-T}Dynamical masses as functions of the temperature
for $\phi$ mesons at $eB=5\,m_{\pi}^{2}$. Masses for bound states
with spin $\lambda=0$ and $\lambda=\pm1$ are plotted with red solid
lines and blue solid lines, respectively, while those for resonance
excitations are plotted with red dashed lines and blue dashed lines,
respectively.}
\end{figure}

\begin{figure}
\includegraphics[width=8cm]{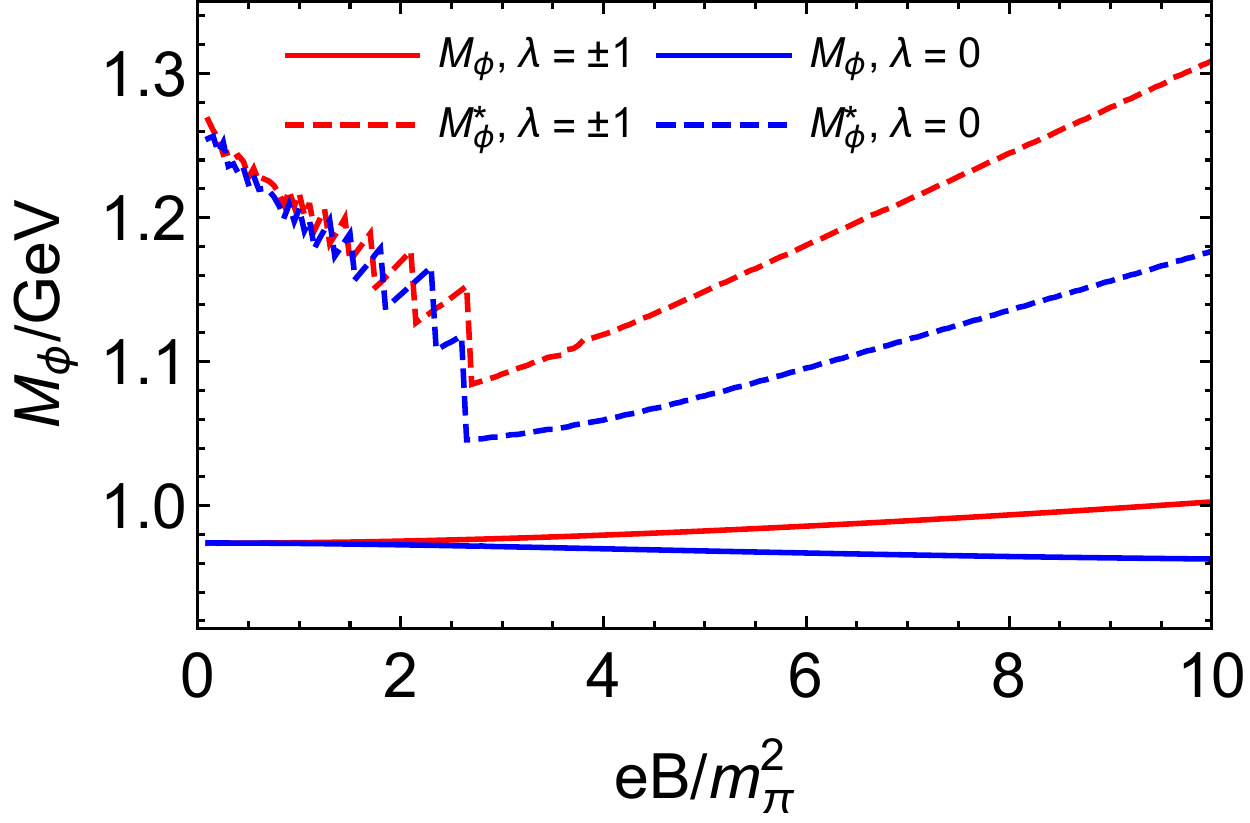}

\caption{\label{fig:Mphi-eB}Dynamical masses as functions of $eB$ for $\phi$
mesons at $T=150$ MeV. Notations are the same as in Fig. \ref{fig:Vector-Mass-T}.}
\end{figure}

Denoting the energy for the most significant peak in the continuum
as $M_{\phi,\lambda}^{\ast}$, we plot $M_{\phi,\lambda}^{\ast}$,
and $M_{\phi,\lambda}$ for bound states, as functions of the
temperature at $eB=5m_{\pi}^{2}$ in Fig. \ref{fig:Vector-Mass-T}.
Similar to the temperature dependence of the $s$ quark, the $\phi$
meson masses also decrease at higher temperatures. One can observe
from Fig. \ref{fig:Vector-Mass-T} the mass splitting between states
with $\lambda=0$ and $\lambda=\pm1$, induced by the broken symmetry
because of the magnetic field. States with $\lambda=0$ always have
smaller masses compared to states with $\lambda=\pm1$. As the temperature
increases, the bound states finally dissociate, and the corresponding
dissociation temperature is $215$ MeV for $\lambda=\pm1$ and $270$
MeV for $\lambda=0$. Therefore when the temperature $215\,\text{MeV}<T<270\,\text{MeV}$,
$\phi$ meson bound states are purely at states with $\lambda=0$.
In Fig. \ref{fig:Mphi-eB}, we show the field strength dependence
for the $\phi$ meson masses.The mass for $\lambda=0$ bound-states
decreases with an increasing $eB$, while that for $\lambda=\pm1$
increases. Masses for resonance excitations oscillate when $eB<2.5\,m_{\pi}^{2}$,
and is nearly linear in $eB$ when $2.5\,m_{\pi}^{2}<eB<10\,m_{\pi}^{2}$. 

\subsection{Spin alignment for $\phi$ mesons}

Substituting the spectral functions calculated in subsection \ref{subsec:Spectra functions}
into Eqs. (\ref{eq:density matrix}) and (\ref{eq:spin alignment}),
we derive the $\phi$ meson's spin alignment. Note that the bound
states and resonance excitations are at different mass regions: the
bound states have masses $\sim1$ GeV or smaller, while masses for
the resonance excitations have larger masses. Therefore we treat them
as different kinds of particles and calculate their spin alignments
separately.

\begin{figure}
\includegraphics[width=8cm]{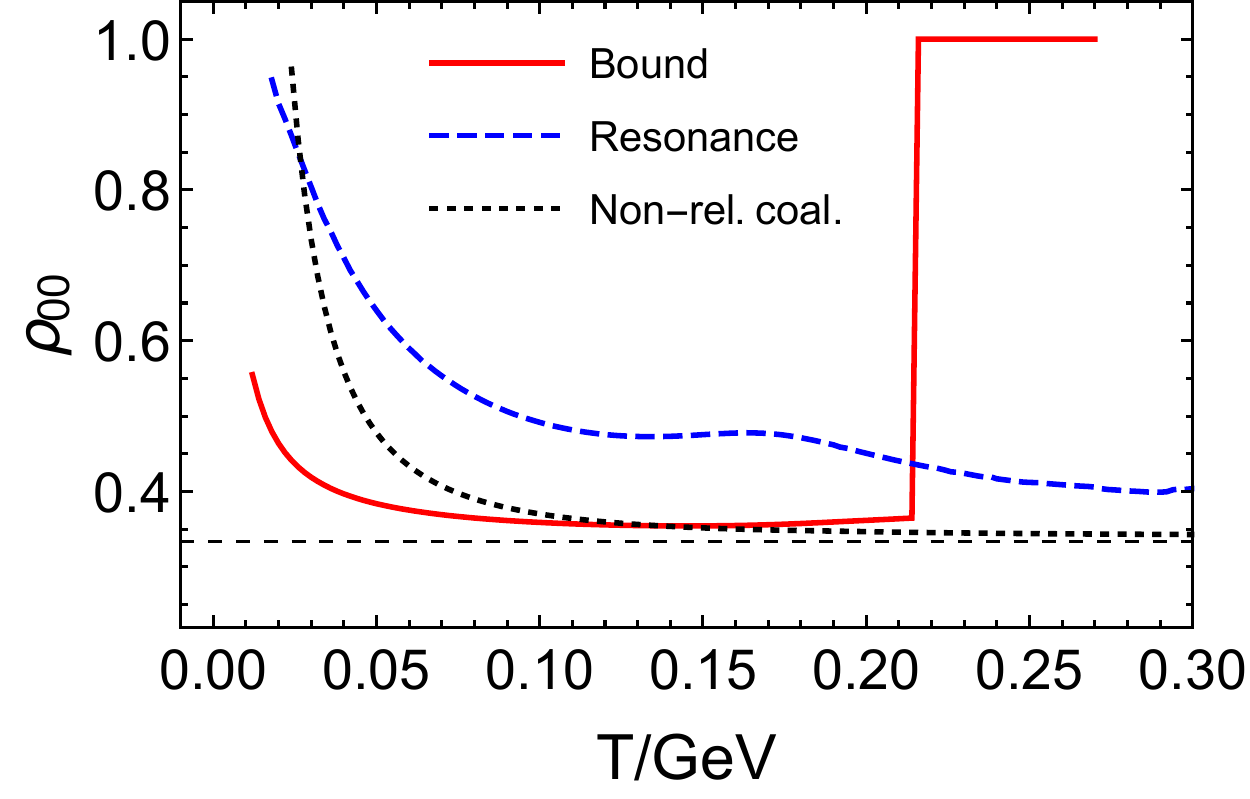}

\caption{\label{fig:Spin-alignment-T}Spin alignments as functions of the temperature.
The magnetic field strength is taken as $eB=5\,m_{\pi}^{2}$. Red solid lines are spin alignment for bound states,
calculated using the delta-function part in the spectral function
Eq. (\ref{eq:spectra-separation}), while blue dashed lines are spin alignment
for resonance excitations, calculated using the continuum part in
Eq. (\ref{eq:spectra-separation}). Black dash-dotted lines are results
from the non-relativistic coalescence model, given in Eq. (\ref{eq:non-rel coal}).}
\end{figure}

\begin{figure}
\includegraphics[width=8cm]{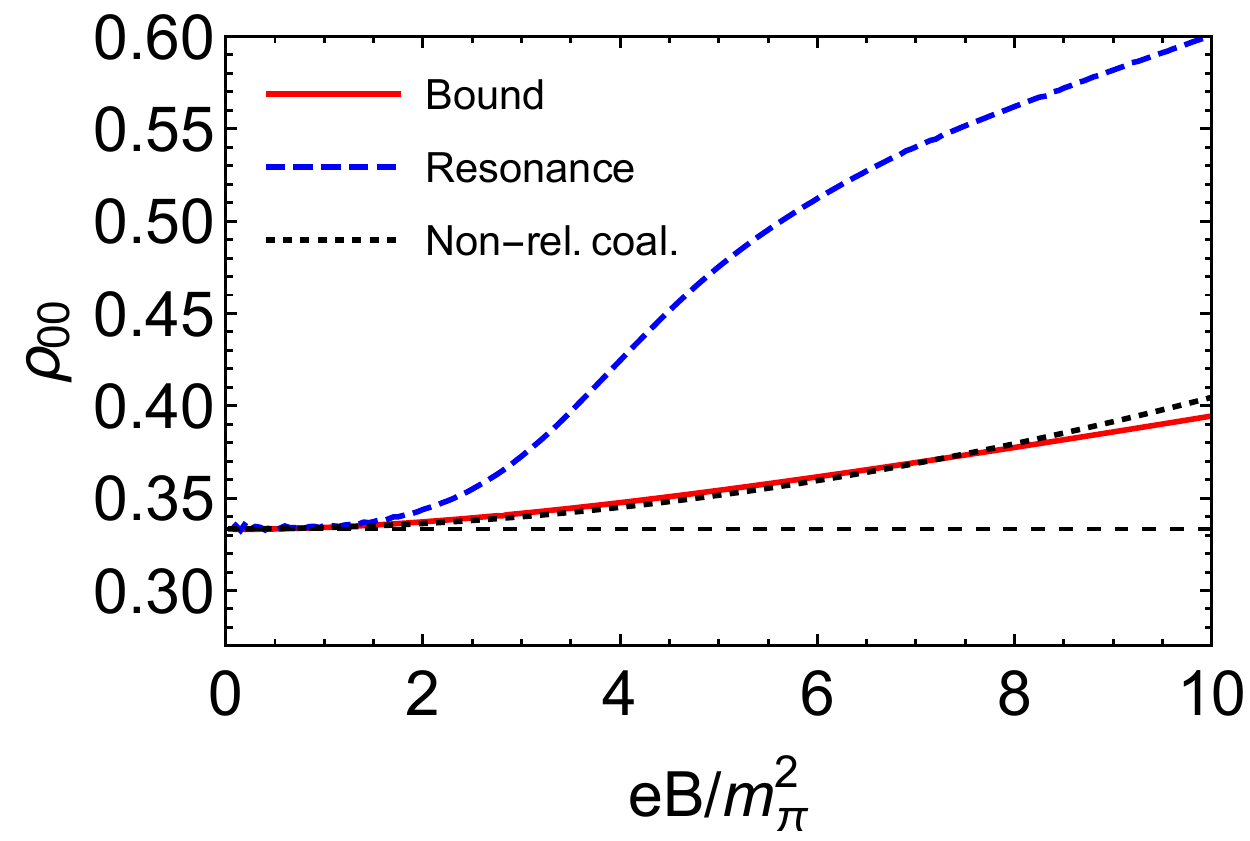}

\caption{\label{fig:Spin-alignment-eB}Spin alignments as functions of the
magnetic field strength at temperature $T=150$ MeV. Notations are
the same as in Fig. \ref{fig:Spin-alignment-T}.}
\end{figure}

In Fig. \ref{fig:Spin-alignment-T}, we plot spin alignments $\rho_{00}$
as functions of the temperature at $eB=5\,m_{\pi}^{2}$. Spin alignments
for bound states and resonance excitations are denoted as red solid
lines and blue dashed lines, respectively. As a comparison, we also
plot the result from a non-relativistic coalescence model \citep{Yang:2017sdk},
\begin{equation}
\rho_{00}^{\text{Non-rel. coal.}}=\frac{1}{3}+\frac{4}{9T^{2}}\mu_{s}^{2}B^{2}\,,\label{eq:non-rel coal}
\end{equation}
where $\mu_{s}=q_{s}/(2M_{s})$ is the magnetic moment for the $s$
quark, with $q_{s}=(-1/3)e$. The spin alignment $\rho_{00}$ for
bound states, shown by the red solid line in Fig. \ref{fig:Spin-alignment-T},
is significantly smaller than the result from the non-relativistic
coalescence model (black dotted line) when $T<100$ MeV. The $\rho_{00}$
for bound states decreases towards $1/3$ with an increasing $T$.
For the temperature $T=150$ MeV, $\rho_{00}$ for bound states is
in very good agreement with the coalescence model. Above 150 MeV,
$\rho_{00}$ for bound states is larger than the result from the coalescence
model. Especially, bound states with $\lambda=\pm1$ vanish when $T>215$
MeV, leading to the result of $\rho_{00}=1$ in this temperature region.
On the other hand, the spin alignment $\rho_{00}$ for resonance states
is always larger than the result from the coalescence model except
at very low temperatures. In Fig. \ref{fig:Spin-alignment-eB} we
show spin alignments as functions of the magnetic field strength.
We focus on a fixed temperature $T=150$ MeV and observe that $\rho_{00}$
increases with increasing $eB$. In the zero-field limit $eB\rightarrow0$,
$\rho_{00}$ agrees with $1/3$, as expected. We find that the spin
alignment for bound states agrees with the result of the non-relativistic
coalescence model, while the spin alignment for resonance excitations
is significantly larger. 

\section{Effect of anomalous magnetic field \label{sec:Effect-of-anomalous}}

Considering that constituent quark may have different magnetic moments
compared with free quarks \citep{Brekke:1987cc,Chang:2010hb,Fayazbakhsh:2014mca,Ayala:2015bgv,Chaudhuri:2019lbw,Xu:2020yag},
we study the effect of AMM in this section. The AMM is included in
the Lagrangian $\mathcal{L}_{q}$ by a term $q_{f}\kappa_{f}F_{\mu\nu}\sigma^{\mu\nu}/2$
as
\begin{equation}
\mathcal{L}_{q}=\sum_{f=u,d,s}\overline{\psi}_{f}\left(i\gamma_{\mu}D_{f}^{\mu}-m_{f}-\frac{1}{2}q_{f}\kappa_{f}F_{\mu\nu}\sigma^{\mu\nu}\right)\psi_{f}\,,
\end{equation}
where $\kappa_{f}$ denote the AMMs for quarks with flavor $f=u,d,s$.
By applying the Foldy--Wouthuysen transformation \cite{Foldy:1949wa},
one can show that the magnetic moment for quark is modified to $\mu_{f}=(1+2M_{f}\kappa_{f})q_{f}/(2M_{f})$,
where $M_{f}$ is the dynamical mass that includes the contribution
of chiral condensate. By fitting the phenomenological values of magnetic
moments for valence quarks \citep{Dothan:1981ex}, i.e., $\mu_{u}=2.08\mu_{N}$,
$\mu_{d}=-1.31\mu_{N}$, and $\mu_{s}=-0.77\mu_{N}$, where the nuclear
magneton $\mu_{N}\equiv e/2m_{p}$ with $m_{p}=0.938\ \text{GeV}$
being the proton mass, we derive the following set of AMMs, 
\begin{align}
\kappa_{u}=0.123\,\text{GeV}^{-1}, & \kappa_{d}=0.555\,\text{GeV}^{-1},\nonumber \\
\kappa_{s}=0.329\,\text{GeV}^{-1}.\label{eq:nonzero AMM}
\end{align}
In later parts of this section, this set of AMM is denoted as $\kappa_{f}\neq0$
for simplicity.

\subsection{Quark mass}

We first focus on dynamical quark masses. In the presence of AMMs,
the dispersion relation for Landau levels reads 
\begin{eqnarray}
E & = & \pm E_{f,n,s}(p_{z})\nonumber \\
 & = & \pm\sqrt{p_{z}^{2}+\left(\sqrt{M_{f}^{2}+2n|q_{f}B|}-s\kappa_{f}|q_{f}B|\right)^{2}}\,,\label{eq:eigenenergies-1}
\end{eqnarray}
where $s=+$ for the lowest Landau level and $s=\pm$ for Landau levels
$n\ge1$ denote the spin state. The AMMs induce an additional spin-magnetic
coupling and therefore Landau levels $n\ge1$ are no-longer two-fold
degenerate in spin. The grand thermodynamic potential is constructed
in a similar way as Eq. (\ref{eq:quark_potential}) and the chiral
condensates are still solved by $\partial\Omega/\partial\sigma_{f}=0$.

\begin{figure}
\includegraphics[width=8cm]{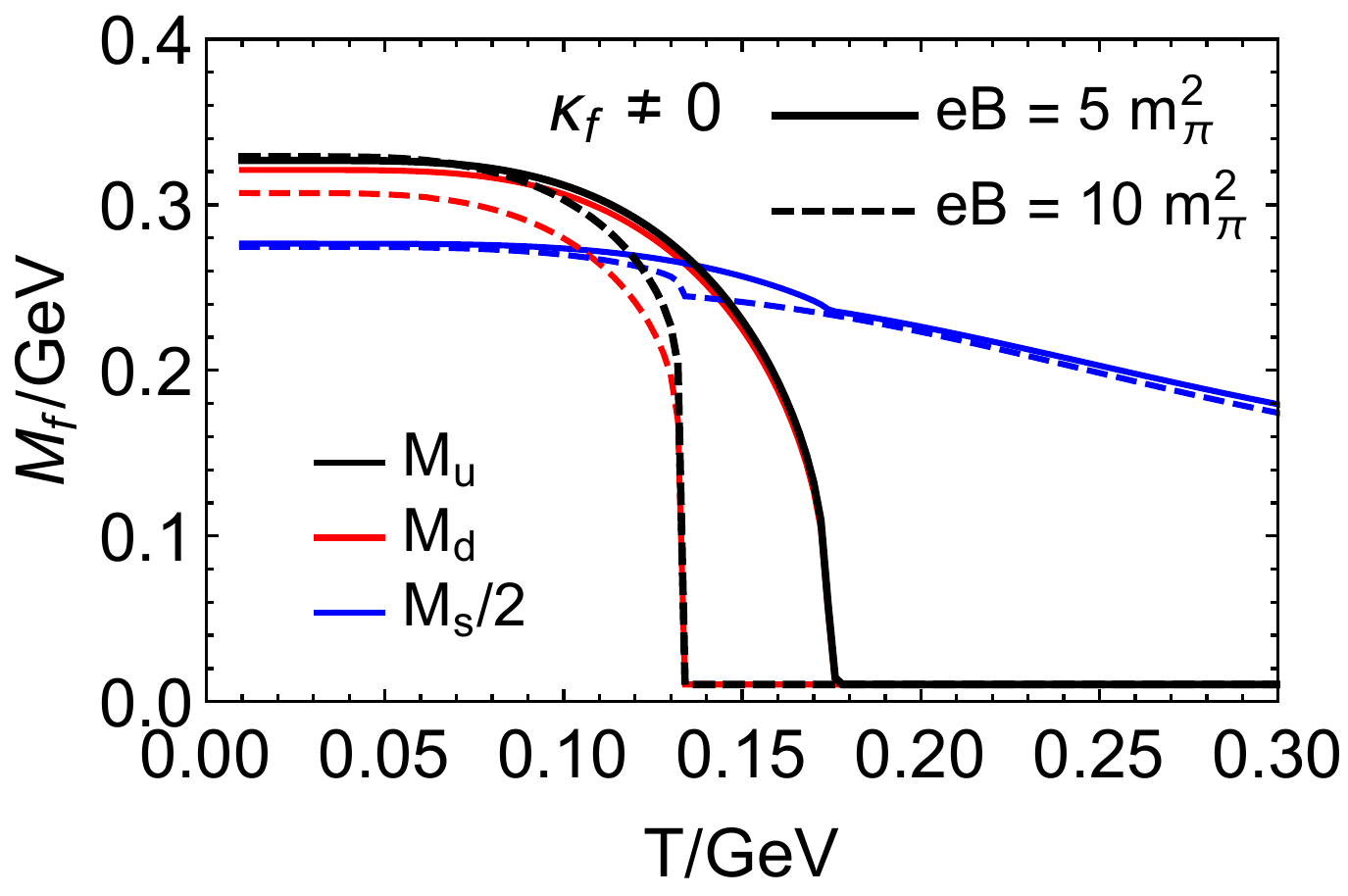}

\caption{\label{fig:Dynamical-masses-1}Dynamical masses as functions of the
temperature for $u$ quark (black lines), $d$ quark (red lines),
and $s$ quark (blue lines) with AMMs $\kappa_{f}\protect\neq0$ in
magnetic fields $eB=5\,m_{\pi}^{2}$ (solid lines) and $eB=10\,m_{\pi}^{2}$
(dashed lines).}
\end{figure}

\begin{figure}
\includegraphics[width=8cm]{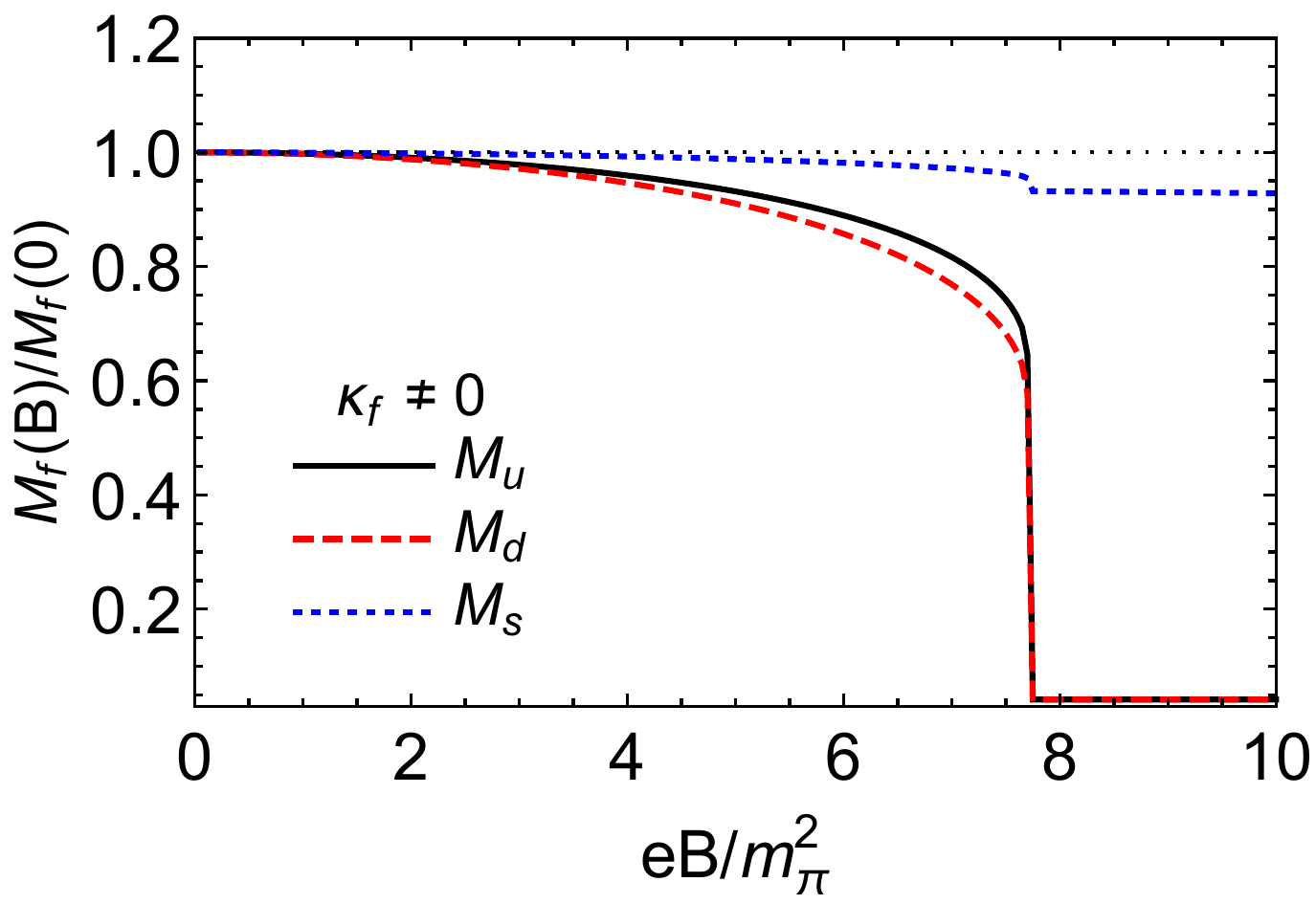}

\caption{\label{fig:Dynamical-mass-eB}Dynamical masses as functions of the
magnetic field strength at $T=150$ MeV for quarks with nonvanishing
AMMs, normalized by masses when $eB=0$. Dynamical masses for $u$,
$d$, and $s$ quarks are plotted with black solid line, red dashed
line, and blue dotted line, respectively.}
\end{figure}

We plot in Fig. \ref{fig:Dynamical-masses-1} dynamical masses as
functions of the temperature for $u$, $d$, and $s$ quarks with
$\kappa_{f}\neq0$. As $T$ increases, masses of $u$ and $d$ quarks
sharply decrease to their current masses, indicating a first-order
phase transition that happens at $176$ MeV when $eB=5\,m_{\pi}^{2}$,
and at $133$ MeV when $eB=10\,m_{\pi}^{2}$. Such a behaviour is
significantly different from the cross-over in Fig. \ref{fig:Dynamical-masses},
indicating that the phase structure for light quarks are strongly
affected by AMMs. On the other hand, the $s$ quark is less affected
by the AMM and still undergoes a cross-over. In the chiral symmetry
breaking phase, the quark masses at $eB=5\,m_{\pi}^{2}$ is larger
than those at $eB=10\,m_{\pi}^{2}$, which is the inverse magnetic
catalysis phenomena. The dependence to the magnetic field strength
is shown in Fig. \ref{fig:Dynamical-mass-eB}. We observe that the
nonzero AMMs result in the inverse magnetic catalysis, i.e., the dynamical
masses decrease with an increasing field strength, which is opposite
to the case with $\kappa_{f}=0$ as shown in Fig. \ref{fig:Dynamical-mass-eB-1}.
Moreover, at a particular field strength, $M_{u}$ and $M_{d}$ suffer
a discontinuity, corresponding to a first-order phase transition.
For the set of AMMs in Eq. (\ref{eq:nonzero AMM}), the critical field
strength is $eB_{C}=7.75\,m_{\pi}^{2}$ at $T=150$ MeV. Above this
$eB_{C}$, dynamical masses for $u$ and $d$ quarks are consistent
with current quark masses.

\subsection{Spectral function for $\phi$ mesons}

\begin{figure}
\includegraphics[width=8cm]{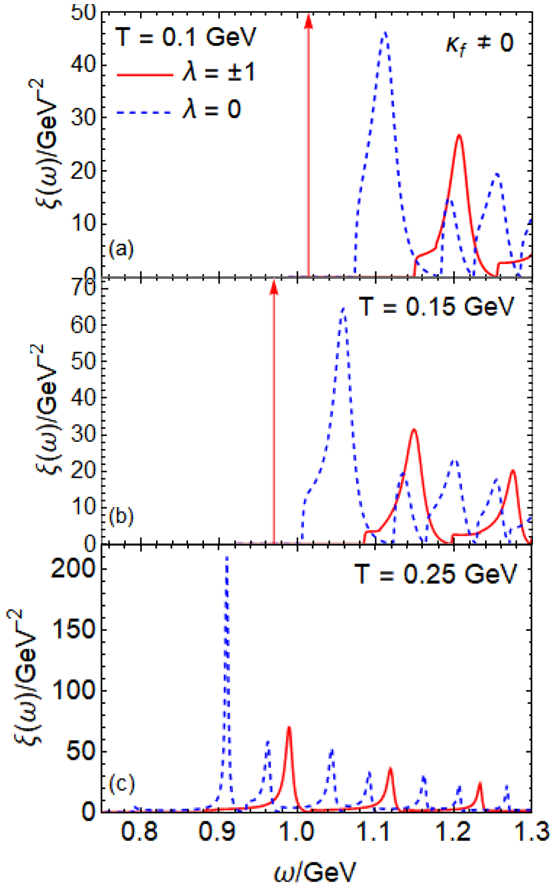}

\caption{\label{fig:Mass-spectra-1}Spectral functions for $\phi$ mesons in
a constant magnetic field with $eB=5m_{\pi}^{2}$, at $T=100$ MeV
{[}panels (a){]}, $T=150$ MeV {[}panels (b){]}, and $T=250$ MeV
{[}panels (c){]}. The AMMs are set to nonzero values given in Eq.
(\ref{eq:nonzero AMM}). Here red solid lines denote spectral functions
for $\phi$ mesons with spin state $\lambda=\pm1$ and blue dashed
lines denote the spectral function of $\lambda=0$. Red and blue arrows
correspond to delta functions.}
\end{figure}

\begin{figure}
\includegraphics[width=8cm]{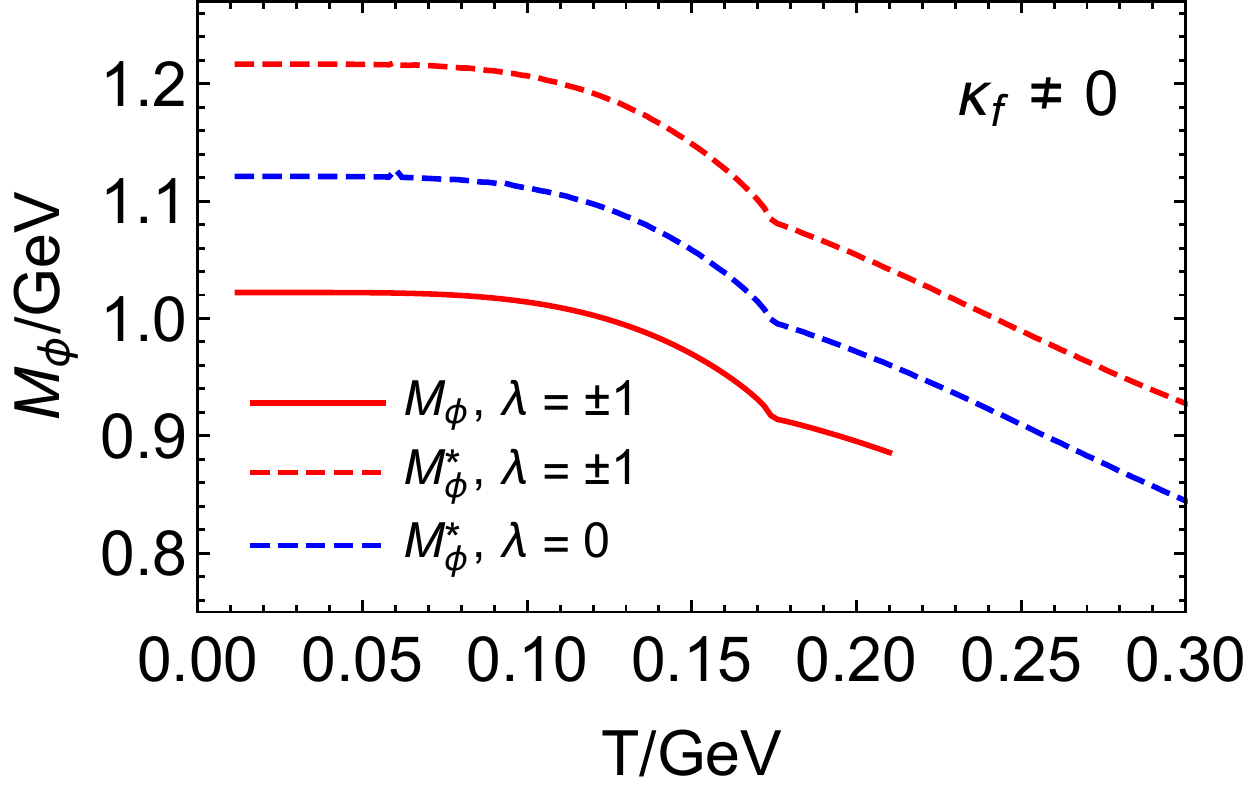}

\caption{\label{fig:Vector-Mass-T-1}Dynamical masses as functions of the temperature
for $\phi$ mesons at $eB=5\,m_{\pi}^{2}$. Masses for bound states
with spin $\lambda=0$ and $\lambda=\pm1$ are plotted with red solid
lines and blue solid lines, respectively, while those for resonance
excitations are plotted with red dashed lines and blue dashed lines,
respectively. The AMMs for quarks are nonzero values given in Eq. (\ref{eq:nonzero AMM}).}
\end{figure}

\begin{figure}
\includegraphics[width=8cm]{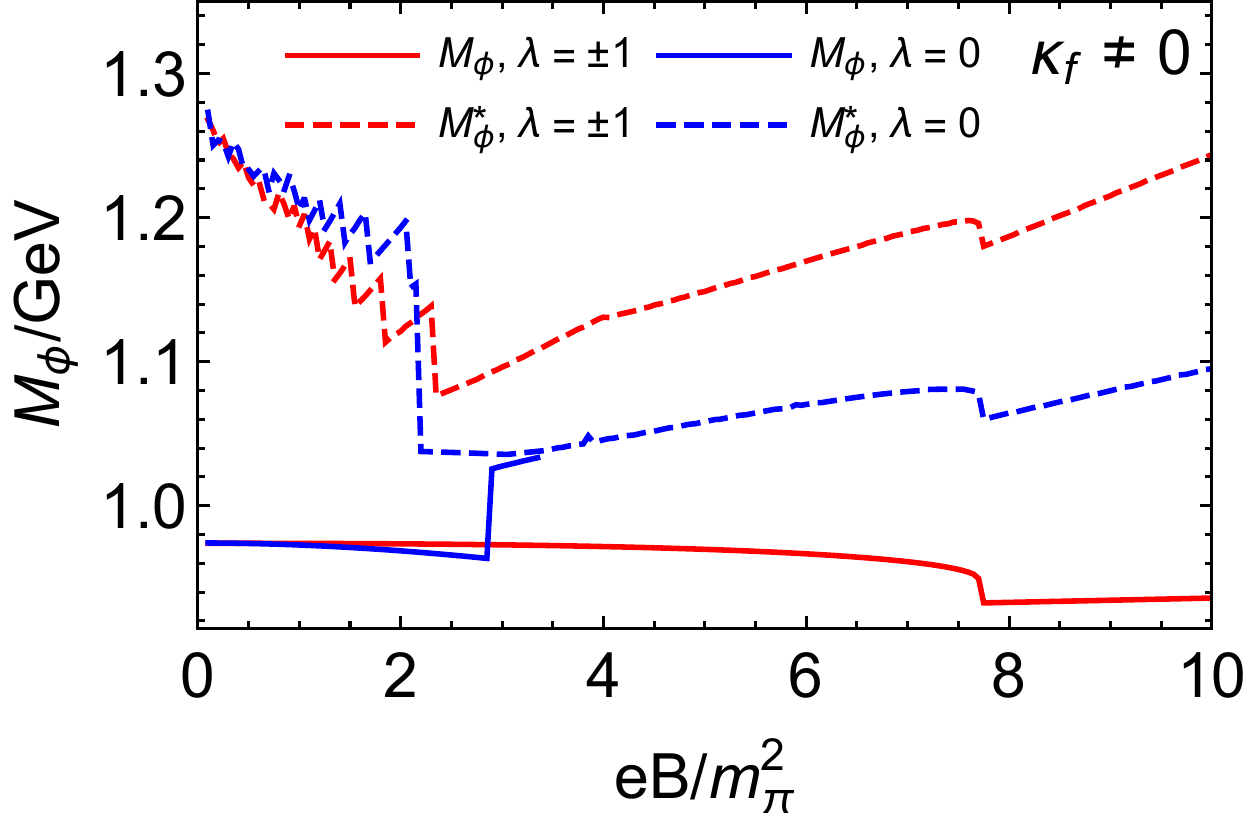}

\caption{\label{fig:Mphi-eB-1}Dynamical masses as functions of $eB$ for $\phi$
mesons at $T=150$ MeV. Notations are the same as in Fig. \ref{fig:Vector-Mass-T}.}
\end{figure}

The propagator for vector meson is derived by solving the Dyson-Schwinger
equation (\ref{eq:Dyson equation}) with self-energy (\ref{eq:self-energy}).
In the presence of AMMs, the quark propagator is given by the following
form, 
\begin{equation}
S_{f}(p)=ie^{-p_{\perp}^{2}/|q_{f}B|}\sum_{n=0}^{\infty}(-1)^{n}D_{n}^{f}(p)\frac{1}{\mathcal{M}(p)-2n|q_{f}B|}\,,\label{eq:fermion propagator-1}
\end{equation}
where $D_{n}^{f}(p)$ is given by Eq. (\ref{eq:Df-n}) with $M_{f}$
replaced by $M_{f}+i\kappa_{f}q_{f}B\gamma^{1}\gamma^{2}$. The explicit
form of the matrix $\mathcal{M}$ is given in Ref. \citep{Miransky:2015ava},
from which we derive 
\begin{align}
\frac{1}{\mathcal{M}(p)-2n|q_{f}B|} & =\frac{1}{U_{n}}\left[(p^{0})^{2}-(p_{z})^{2}-M_{f}^{2}-2n|q_{f}B|\right.\nonumber \\
 & \left.\!\!\!\!\!\!\!\!\!\!\!\!\!\!+(\kappa_{f}q_{f}B)^{2}+2\kappa_{f}q_{f}B(p_{z}\gamma^{0}\gamma^{5}-p^{0}\gamma^{3}\gamma^{5})\right]\,,
\end{align}
where the denominator $U_{n}$ is given by 
\begin{eqnarray}
U_{n} & = & \left[(p^{0})^{2}-(p_{z})^{2}-M_{f}^{2}-2n|q_{f}B|-(\kappa_{f}q_{f}B)^{2}\right]^{2}\nonumber \\
 &  & -4(M_{f}^{2}+2n|q_{f}B|)(\kappa_{f}q_{f}B)^{2}\,.
\end{eqnarray}
We note that the eigenenergies in Eq. (\ref{eq:eigenenergies-1})
are determined by solutions of $U_{n}=0$. One can also prove that
the propagator reduce to Eq. (\ref{eq:fermion propagator}) in the
absence of $\kappa_{f}$. Using the quark propagator and the dynamical
mass obtained in the previous subsection, we are then able to calculate
the vector meson's spectral function. We still focus on the $\phi$
meson at static, ${\bf k={\bf 0}}$. The spectral functions at $T=100$
MeV, $150$ MeV, and 200 MeV are shown in Fig. \ref{fig:Mass-spectra-1}.
The magnetic field strength is set to $eB=5\,m_{\pi}^{2}$, leading
to the difference between longitudinally $(\lambda=0)$ and transversely
$(\lambda=\pm1)$ polarized states. We also observe the bound state
with $\lambda=0$ vanishes for all three considered temperatures,
while bound states with $\lambda=\pm1$ vanish at $T=250$ MeV and
exist at $T=100$ MeV or $150$ MeV. The temperature dependence for
the peak masses are shown in Fig. (\ref{fig:Vector-Mass-T-1}), where
we observe that the bound state with $\lambda=0$ does not exist even
at lower temperatures. Compared Fig. \ref{fig:Vector-Mass-T-1} with
Fig. \ref{fig:Vector-Mass-T}, we find that the $\lambda=\pm1$ states
and the resonance masses are nearly not affected by the AMMs. We also
plot in Fig. \ref{fig:Mphi-eB-1} the $\phi$ meson masses as functions of magnetic field strength. When $eB<2.9\,m_{\pi}^{2}$, masses for bound states
decrease in larger magnetic fields. The bound state with $\lambda=0$
suddenly increases approach the mass of the resonance state at $eB\sim2.9\,m_{\pi}^{2}$
and then dissociate when $eB\gtrsim3.4\,m_{\pi}^{2}$, corresponding
to the Mott transition. The masses for resonance excitations are nearly
independent to the AMMs, as compared to Fig. \ref{fig:Mphi-eB}.

\subsection{Spin alignment for $\phi$ mesons}

\begin{figure}
\includegraphics[width=8cm]{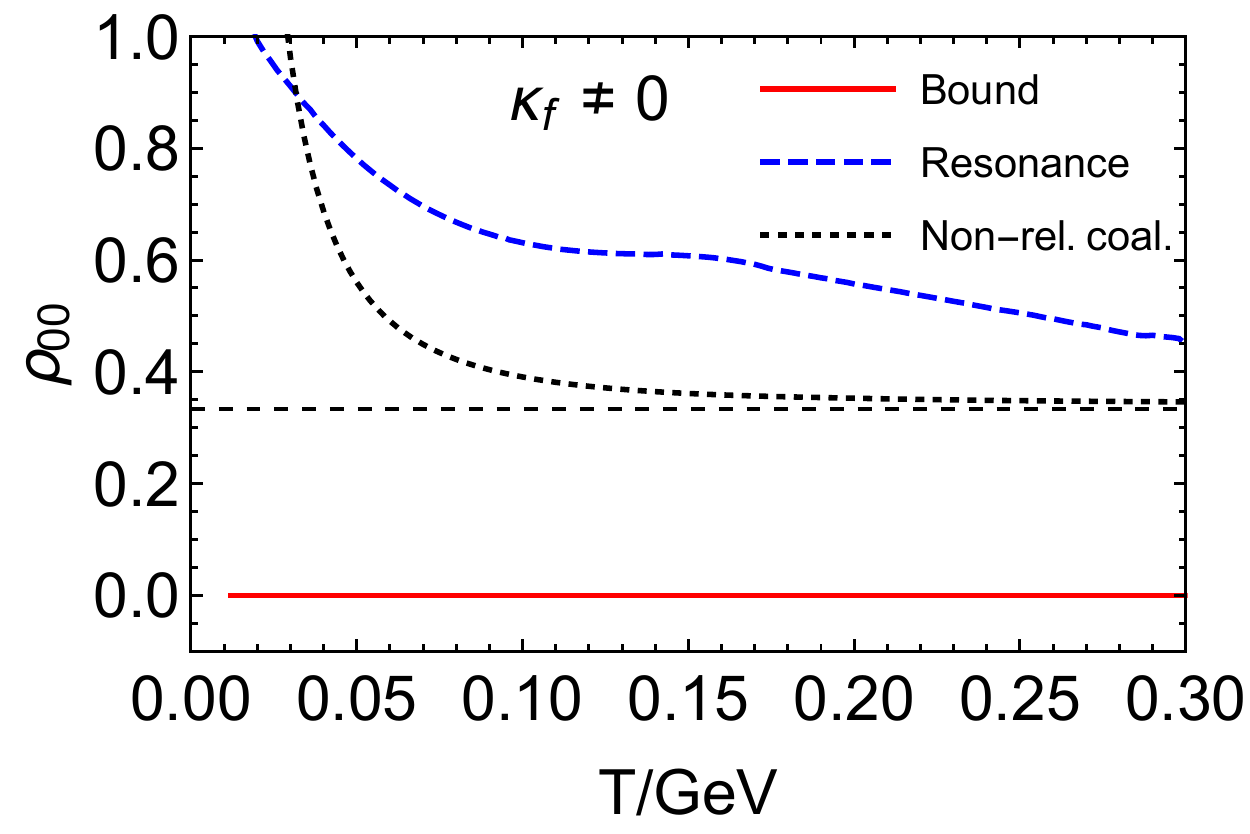}

\caption{\label{fig:Spin-alignment-T-1}Spin alignments as functions of the
temperature. The magnetic field strength is taken as $eB=5\,m_{\pi}^{2}$.
The quark AMMs are set to nonzero values given in Eq. (\ref{eq:nonzero AMM}).}
\end{figure}

\begin{figure}
\includegraphics[width=8cm]{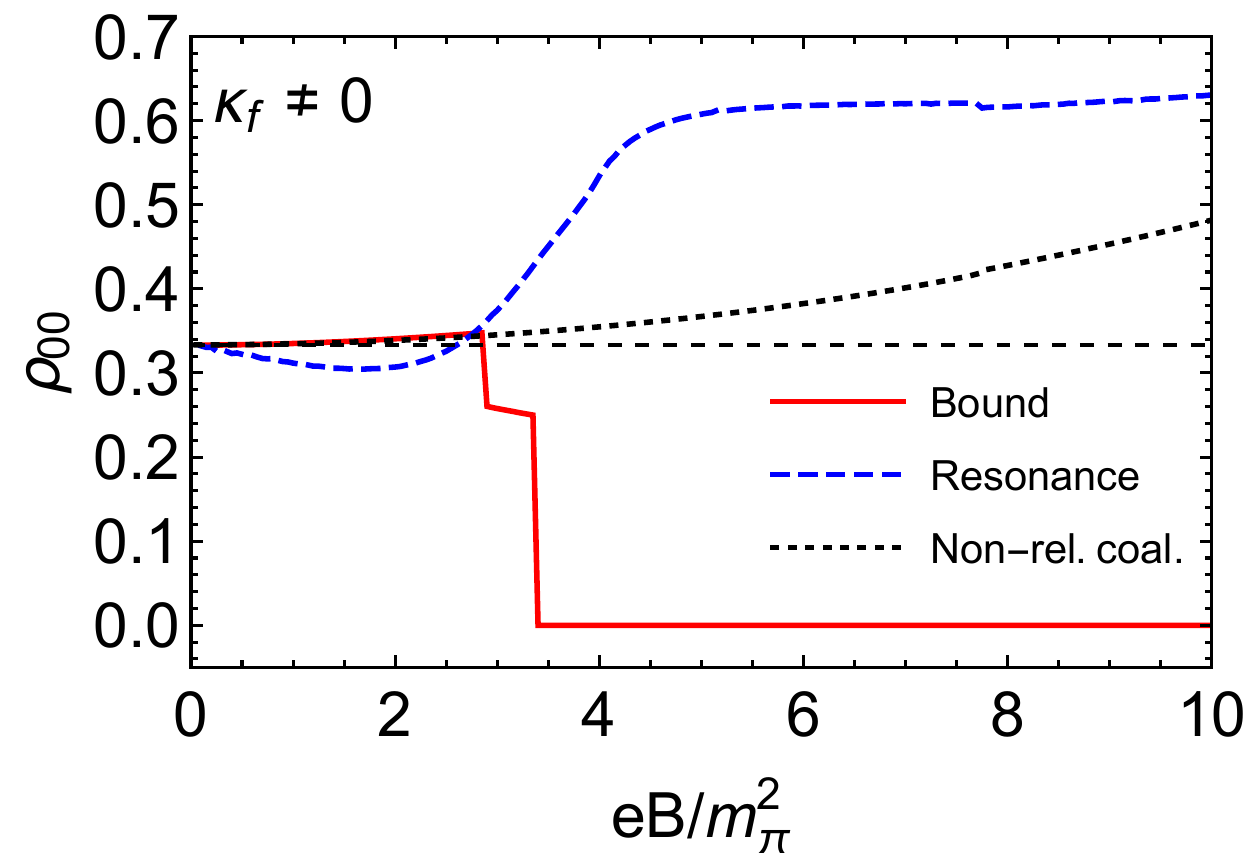}

\caption{\label{fig:Spin-alignment-eB-1}Spin alignments as functions of the
magnetic field strength at temperature $T=150$ MeV. Notations are
the same as in Fig. \ref{fig:Spin-alignment-T}.}
\end{figure}

Substituting the spectral functions into Eqs. (\ref{eq:density matrix})
and (\ref{eq:spin alignment}), we derive the $\phi$ meson's spin
alignment. For the case with nonzero AMMs, the spin alignments as
functions of temperature and magnetic field strength are shown in
Figs. \ref{fig:Spin-alignment-T-1} and \ref{fig:Spin-alignment-eB-1},
respectively. As a comparison, we also show results from a non-relativistic
coalescence model \citep{Yang:2017sdk}, given by Eq. (\ref{eq:non-rel coal}),
with the magnetic moment $\mu_{s}=(1+2M_{s}\kappa_{s})q_{s}/(2M_{s})$
including the effect of AMM. From Fig. \ref{fig:Spin-alignment-T-1}
we observe that the spin alignment for bound states is always equal
to zero because all bound states have $\lambda=\pm1$, as indicated
by Fig. \ref{fig:Vector-Mass-T-1}. The resonance states still have
$\rho_{00}>\rho_{00}^{\text{Non-rel. coal.}}>1/3$, which is similar
to the case with $\kappa_{f}=0$. On the other hand, in a weak magnetic
field with $eB<2.5\,m_{\pi}^{2}$, $\rho_{00}$ for bound states still
agrees well with the non-relativistic model. It jumps to a negative
value at $eB\sim2.9\,m_{\pi}^{2}$, which is a straightforward result
of the Mott transition for states with $\lambda=0$. When the magnetic
field strength $eB\gtrsim3.4\,m_{\pi}^{2}$, the bound state with
$\lambda=0$ dissociates in the thermal medium, resulting in a vanishing
spin alignment $\rho_{00}=0$. Meanwhile, the $\rho_{00}$ for resonance
excitations shows a non-monotonic structure: $\rho_{00}<1/3$ when
$eB<2.6\,m_{\pi}^{2}$ and $\rho_{00}>1/3$ when $eB>2.6\,m_{\pi}^{2}$.

\section{Summary \label{sec:Summary}}

In this manuscript, we study the mass splitting and the spin alignment
for the vector $\phi$ meson in a hot magnetized matter. The three-flavor
NJL model is used to properly include the chiral phase transition.
In this framework, mesons are described as quantum fluctuations beyond
the mean-field background. The meson's propagator is given by resuming
quark bubbles at the random phase approximation. For a vector meson,
the propagator is a Lorentz tensor perpendicular to the meson's momentum,
which can be further projected into the spin space. The spin density
matrix is then derived by the convolution of the spectral function
and the Bose-Einstein distribution, while the spectral function is
given by the imaginary part of the propagator. In this way, we are
able to calculate the vector meson's spin alignment  in a thermal medium.

If we choose the spin quantization direction for the vector meson
as the direction of the external magnetic field, the density matrix
is then diagonal and spin states $\lambda=\pm1$ are degenerate. We
numerically calculated the spectral functions for $\phi$ mesons.
In general the spectral function can be separated into a delta-function
part corresponding to a quark-antiquark bound state, and a continuum
part corresponding to resonance excitations. We observe multi-peak
structures in the continuum part, which is the result of Landau quantization
for constituent quarks. The magnetic field leads to a mass splitting
between states with $\lambda=0$ and $\lambda=\pm1$, resulting in
the spin alignment $\rho_{00}\neq1/3$ for the $\phi$ meson in a
hot medium. If we only focus on the bound states, we find that $\rho_{00}>1/3$
and is in good agreement with the result from the non-relativistic
coalescence model. However, the spin alignment for resonance states
gives a much larger result.

Since constituent quarks and free quarks have different magnetic moments,
we incorporate the quark AMMs according to the constituent quark magnetic
moments observed in experiments. The light-flavor quarks then show
the inverse magnetic catalysis behaviors, and the chiral phase transition
is first order. The AMMs then significantly modify the mass spectra
and the spin alignment for the $\phi$ meson, especially in a strong
magnetic field. Considering a different
choice of AMMs, like in Ref. \citep{Kawaguchi:2022dbq}, may result
in a different $\rho_{00}$, which is waiting for more comprehensive
studies in the future.

Even though the spin quantization direction is fixed to the direction
of the magnetic field in our numerical calculations, we can easily
generalize the results to the case that $\rho_{00}$ is measured along
any other direction, c.f., Eq. (\ref{eq:general result}). This is
achieved by performing a rotation in spin space for the density matrix.
Therefore results in this work can be applied to study the $\phi$
meson's spin alignment in a fluctuating magnetic field. When the fluctuation
is anisotropic in space, the spin alignment $\rho_{00}$ will deviate
from $1/3$. Therefore this work may help us better understand the
role of the magnetic field in the $\phi$ mesons' spin alignment. 
\begin{acknowledgments}
The authors thank  Mei Huang and Hai-Cang  Ren for enlightening discussions. This work is supported in part by the National Key Research and Development Program of China under Contract No. 2022YFA1604900.  X.L.S.
is supported by the National Natural Science Foundation of China (NSFC)
under grant No. 12047528, and by the Project funded by China Postdoctoral
Science Foundation under grant No. 2021M701369. S.Y.Y, Y.L.Z., and
D.F.H. are supported by the National Natural Science Foundation of
China (NSFC) under grant Nos. 12275104 11890711, and 11890710. 
\end{acknowledgments}

\bibliographystyle{apsrev}
\bibliography{Mass_splitting_and_spin_alignment}

\begin{thebibliography}{87}
\expandafter\ifx\csname natexlab\endcsname\relax\def\natexlab#1{#1}\fi
\expandafter\ifx\csname bibnamefont\endcsname\relax
  \def\bibnamefont#1{#1}\fi
\expandafter\ifx\csname bibfnamefont\endcsname\relax
  \def\bibfnamefont#1{#1}\fi
\expandafter\ifx\csname citenamefont\endcsname\relax
  \def\citenamefont#1{#1}\fi
\expandafter\ifx\csname url\endcsname\relax
  \def\url#1{\texttt{#1}}\fi
\expandafter\ifx\csname urlprefix\endcsname\relax\def\urlprefix{URL }\fi
\providecommand{\bibinfo}[2]{#2}
\providecommand{\eprint}[2][]{\url{#2}}

\bibitem[{\citenamefont{Rafelski and Muller}(1976)}]{Rafelski:1975rf}
\bibinfo{author}{\bibfnamefont{J.}~\bibnamefont{Rafelski}} \bibnamefont{and}
  \bibinfo{author}{\bibfnamefont{B.}~\bibnamefont{Muller}},
  \bibinfo{journal}{Phys. Rev. Lett.} \textbf{\bibinfo{volume}{36}},
  \bibinfo{pages}{517} (\bibinfo{year}{1976}).

\bibitem[{\citenamefont{Kharzeev et~al.}(2008)\citenamefont{Kharzeev, McLerran,
  and Warringa}}]{Kharzeev:2007jp}
\bibinfo{author}{\bibfnamefont{D.~E.} \bibnamefont{Kharzeev}},
  \bibinfo{author}{\bibfnamefont{L.~D.} \bibnamefont{McLerran}},
  \bibnamefont{and} \bibinfo{author}{\bibfnamefont{H.~J.}
  \bibnamefont{Warringa}}, \bibinfo{journal}{Nucl. Phys. A}
  \textbf{\bibinfo{volume}{803}}, \bibinfo{pages}{227} (\bibinfo{year}{2008}),
  \eprint{0711.0950}.

\bibitem[{\citenamefont{Skokov et~al.}(2009)\citenamefont{Skokov, Illarionov,
  and Toneev}}]{Skokov:2009qp}
\bibinfo{author}{\bibfnamefont{V.}~\bibnamefont{Skokov}},
  \bibinfo{author}{\bibfnamefont{A.~Y.} \bibnamefont{Illarionov}},
  \bibnamefont{and} \bibinfo{author}{\bibfnamefont{V.}~\bibnamefont{Toneev}},
  \bibinfo{journal}{Int. J. Mod. Phys. A} \textbf{\bibinfo{volume}{24}},
  \bibinfo{pages}{5925} (\bibinfo{year}{2009}), \eprint{0907.1396}.

\bibitem[{\citenamefont{Kharzeev et~al.}(2013)\citenamefont{Kharzeev,
  Landsteiner, Schmitt, and Yee}}]{Kharzeev:2013jha}
\bibinfo{editor}{\bibfnamefont{D.}~\bibnamefont{Kharzeev}},
  \bibinfo{editor}{\bibfnamefont{K.}~\bibnamefont{Landsteiner}},
  \bibinfo{editor}{\bibfnamefont{A.}~\bibnamefont{Schmitt}}, \bibnamefont{and}
  \bibinfo{editor}{\bibfnamefont{H.-U.} \bibnamefont{Yee}}, eds.,
  \emph{\bibinfo{title}{{Strongly Interacting Matter in Magnetic Fields}}},
  vol. \bibinfo{volume}{871} (\bibinfo{year}{2013}), ISBN
  \bibinfo{isbn}{978-3-642-37304-6, 978-3-642-37305-3}.

\bibitem[{\citenamefont{Deng and Huang}(2012)}]{Deng:2012pc}
\bibinfo{author}{\bibfnamefont{W.-T.} \bibnamefont{Deng}} \bibnamefont{and}
  \bibinfo{author}{\bibfnamefont{X.-G.} \bibnamefont{Huang}},
  \bibinfo{journal}{Phys. Rev. C} \textbf{\bibinfo{volume}{85}},
  \bibinfo{pages}{044907} (\bibinfo{year}{2012}), \eprint{1201.5108}.

\bibitem[{\citenamefont{Tuchin}(2013)}]{Tuchin:2013apa}
\bibinfo{author}{\bibfnamefont{K.}~\bibnamefont{Tuchin}},
  \bibinfo{journal}{Phys. Rev. C} \textbf{\bibinfo{volume}{88}},
  \bibinfo{pages}{024911} (\bibinfo{year}{2013}), \eprint{1305.5806}.

\bibitem[{\citenamefont{McLerran and Skokov}(2014)}]{McLerran:2013hla}
\bibinfo{author}{\bibfnamefont{L.}~\bibnamefont{McLerran}} \bibnamefont{and}
  \bibinfo{author}{\bibfnamefont{V.}~\bibnamefont{Skokov}},
  \bibinfo{journal}{Nucl. Phys. A} \textbf{\bibinfo{volume}{929}},
  \bibinfo{pages}{184} (\bibinfo{year}{2014}), \eprint{1305.0774}.

\bibitem[{\citenamefont{Gursoy et~al.}(2014)\citenamefont{Gursoy, Kharzeev, and
  Rajagopal}}]{Gursoy:2014aka}
\bibinfo{author}{\bibfnamefont{U.}~\bibnamefont{Gursoy}},
  \bibinfo{author}{\bibfnamefont{D.}~\bibnamefont{Kharzeev}}, \bibnamefont{and}
  \bibinfo{author}{\bibfnamefont{K.}~\bibnamefont{Rajagopal}},
  \bibinfo{journal}{Phys. Rev. C} \textbf{\bibinfo{volume}{89}},
  \bibinfo{pages}{054905} (\bibinfo{year}{2014}), \eprint{1401.3805}.

\bibitem[{\citenamefont{Tuchin}(2015)}]{Tuchin:2014iua}
\bibinfo{author}{\bibfnamefont{K.}~\bibnamefont{Tuchin}},
  \bibinfo{journal}{Phys. Rev. C} \textbf{\bibinfo{volume}{91}},
  \bibinfo{pages}{064902} (\bibinfo{year}{2015}), \eprint{1411.1363}.

\bibitem[{\citenamefont{Li et~al.}(2016)\citenamefont{Li, Sheng, and
  Wang}}]{Li:2016tel}
\bibinfo{author}{\bibfnamefont{H.}~\bibnamefont{Li}},
  \bibinfo{author}{\bibfnamefont{X.-l.} \bibnamefont{Sheng}}, \bibnamefont{and}
  \bibinfo{author}{\bibfnamefont{Q.}~\bibnamefont{Wang}},
  \bibinfo{journal}{Phys. Rev. C} \textbf{\bibinfo{volume}{94}},
  \bibinfo{pages}{044903} (\bibinfo{year}{2016}), \eprint{1602.02223}.

\bibitem[{\citenamefont{Chen et~al.}(2021)\citenamefont{Chen, Sheng, and
  Ma}}]{Chen:2021nxs}
\bibinfo{author}{\bibfnamefont{Y.}~\bibnamefont{Chen}},
  \bibinfo{author}{\bibfnamefont{X.-L.} \bibnamefont{Sheng}}, \bibnamefont{and}
  \bibinfo{author}{\bibfnamefont{G.-L.} \bibnamefont{Ma}},
  \bibinfo{journal}{Nucl. Phys. A} \textbf{\bibinfo{volume}{1011}},
  \bibinfo{pages}{122199} (\bibinfo{year}{2021}), \eprint{2101.09845}.

\bibitem[{\citenamefont{Yan and Huang}(2021)}]{Yan:2021zjc}
\bibinfo{author}{\bibfnamefont{L.}~\bibnamefont{Yan}} \bibnamefont{and}
  \bibinfo{author}{\bibfnamefont{X.-G.} \bibnamefont{Huang}}
  (\bibinfo{year}{2021}), \eprint{2104.00831}.

\bibitem[{\citenamefont{Wang et~al.}(2022)\citenamefont{Wang, Zhao, Greiner,
  Xu, and Zhuang}}]{Wang:2021oqq}
\bibinfo{author}{\bibfnamefont{Z.}~\bibnamefont{Wang}},
  \bibinfo{author}{\bibfnamefont{J.}~\bibnamefont{Zhao}},
  \bibinfo{author}{\bibfnamefont{C.}~\bibnamefont{Greiner}},
  \bibinfo{author}{\bibfnamefont{Z.}~\bibnamefont{Xu}}, \bibnamefont{and}
  \bibinfo{author}{\bibfnamefont{P.}~\bibnamefont{Zhuang}},
  \bibinfo{journal}{Phys. Rev. C} \textbf{\bibinfo{volume}{105}},
  \bibinfo{pages}{L041901} (\bibinfo{year}{2022}), \eprint{2110.14302}.

\bibitem[{\citenamefont{Fukushima et~al.}(2008)\citenamefont{Fukushima,
  Kharzeev, and Warringa}}]{Fukushima:2008xe}
\bibinfo{author}{\bibfnamefont{K.}~\bibnamefont{Fukushima}},
  \bibinfo{author}{\bibfnamefont{D.~E.} \bibnamefont{Kharzeev}},
  \bibnamefont{and} \bibinfo{author}{\bibfnamefont{H.~J.}
  \bibnamefont{Warringa}}, \bibinfo{journal}{Phys. Rev. D}
  \textbf{\bibinfo{volume}{78}}, \bibinfo{pages}{074033}
  (\bibinfo{year}{2008}), \eprint{0808.3382}.

\bibitem[{\citenamefont{Son and Surowka}(2009)}]{Son:2009tf}
\bibinfo{author}{\bibfnamefont{D.~T.} \bibnamefont{Son}} \bibnamefont{and}
  \bibinfo{author}{\bibfnamefont{P.}~\bibnamefont{Surowka}},
  \bibinfo{journal}{Phys. Rev. Lett.} \textbf{\bibinfo{volume}{103}},
  \bibinfo{pages}{191601} (\bibinfo{year}{2009}), \eprint{0906.5044}.

\bibitem[{\citenamefont{Liang and Wang}(2005{\natexlab{a}})}]{Liang:2004ph}
\bibinfo{author}{\bibfnamefont{Z.-T.} \bibnamefont{Liang}} \bibnamefont{and}
  \bibinfo{author}{\bibfnamefont{X.-N.} \bibnamefont{Wang}},
  \bibinfo{journal}{Phys. Rev. Lett.} \textbf{\bibinfo{volume}{94}},
  \bibinfo{pages}{102301} (\bibinfo{year}{2005}{\natexlab{a}}),
  \bibinfo{note}{[Erratum: Phys.Rev.Lett. 96, 039901 (2006)]},
  \eprint{nucl-th/0410079}.

\bibitem[{\citenamefont{Becattini et~al.}(2017)\citenamefont{Becattini,
  Karpenko, Lisa, Upsal, and Voloshin}}]{Becattini:2016gvu}
\bibinfo{author}{\bibfnamefont{F.}~\bibnamefont{Becattini}},
  \bibinfo{author}{\bibfnamefont{I.}~\bibnamefont{Karpenko}},
  \bibinfo{author}{\bibfnamefont{M.}~\bibnamefont{Lisa}},
  \bibinfo{author}{\bibfnamefont{I.}~\bibnamefont{Upsal}}, \bibnamefont{and}
  \bibinfo{author}{\bibfnamefont{S.}~\bibnamefont{Voloshin}},
  \bibinfo{journal}{Phys. Rev. C} \textbf{\bibinfo{volume}{95}},
  \bibinfo{pages}{054902} (\bibinfo{year}{2017}), \eprint{1610.02506}.

\bibitem[{\citenamefont{Adamczyk et~al.}(2017)}]{STAR:2017ckg}
\bibinfo{author}{\bibfnamefont{L.}~\bibnamefont{Adamczyk}} \bibnamefont{et~al.}
  (\bibinfo{collaboration}{STAR}), \bibinfo{journal}{Nature}
  \textbf{\bibinfo{volume}{548}}, \bibinfo{pages}{62} (\bibinfo{year}{2017}),
  \eprint{1701.06657}.

\bibitem[{\citenamefont{Das et~al.}(2017)\citenamefont{Das, Plumari,
  Chatterjee, Alam, Scardina, and Greco}}]{Das:2016cwd}
\bibinfo{author}{\bibfnamefont{S.~K.} \bibnamefont{Das}},
  \bibinfo{author}{\bibfnamefont{S.}~\bibnamefont{Plumari}},
  \bibinfo{author}{\bibfnamefont{S.}~\bibnamefont{Chatterjee}},
  \bibinfo{author}{\bibfnamefont{J.}~\bibnamefont{Alam}},
  \bibinfo{author}{\bibfnamefont{F.}~\bibnamefont{Scardina}}, \bibnamefont{and}
  \bibinfo{author}{\bibfnamefont{V.}~\bibnamefont{Greco}},
  \bibinfo{journal}{Phys. Lett. B} \textbf{\bibinfo{volume}{768}},
  \bibinfo{pages}{260} (\bibinfo{year}{2017}), \eprint{1608.02231}.

\bibitem[{\citenamefont{G\"ursoy et~al.}(2018)\citenamefont{G\"ursoy, Kharzeev,
  Marcus, Rajagopal, and Shen}}]{Gursoy:2018yai}
\bibinfo{author}{\bibfnamefont{U.}~\bibnamefont{G\"ursoy}},
  \bibinfo{author}{\bibfnamefont{D.}~\bibnamefont{Kharzeev}},
  \bibinfo{author}{\bibfnamefont{E.}~\bibnamefont{Marcus}},
  \bibinfo{author}{\bibfnamefont{K.}~\bibnamefont{Rajagopal}},
  \bibnamefont{and} \bibinfo{author}{\bibfnamefont{C.}~\bibnamefont{Shen}},
  \bibinfo{journal}{Phys. Rev. C} \textbf{\bibinfo{volume}{98}},
  \bibinfo{pages}{055201} (\bibinfo{year}{2018}), \eprint{1806.05288}.

\bibitem[{\citenamefont{Dubla et~al.}(2020)\citenamefont{Dubla, G\"ursoy, and
  Snellings}}]{Dubla:2020bdz}
\bibinfo{author}{\bibfnamefont{A.}~\bibnamefont{Dubla}},
  \bibinfo{author}{\bibfnamefont{U.}~\bibnamefont{G\"ursoy}}, \bibnamefont{and}
  \bibinfo{author}{\bibfnamefont{R.}~\bibnamefont{Snellings}},
  \bibinfo{journal}{Mod. Phys. Lett. A} \textbf{\bibinfo{volume}{35}},
  \bibinfo{pages}{2050324} (\bibinfo{year}{2020}), \eprint{2009.09727}.

\bibitem[{\citenamefont{Zhang et~al.}(2022)\citenamefont{Zhang, Sheng, Pu,
  Chen, Peng, Wang, and Wang}}]{Zhang:2022lje}
\bibinfo{author}{\bibfnamefont{J.-J.} \bibnamefont{Zhang}},
  \bibinfo{author}{\bibfnamefont{X.-L.} \bibnamefont{Sheng}},
  \bibinfo{author}{\bibfnamefont{S.}~\bibnamefont{Pu}},
  \bibinfo{author}{\bibfnamefont{J.-N.} \bibnamefont{Chen}},
  \bibinfo{author}{\bibfnamefont{G.-L.} \bibnamefont{Peng}},
  \bibinfo{author}{\bibfnamefont{J.-G.} \bibnamefont{Wang}}, \bibnamefont{and}
  \bibinfo{author}{\bibfnamefont{Q.}~\bibnamefont{Wang}}
  (\bibinfo{year}{2022}), \eprint{2201.06171}.

\bibitem[{\citenamefont{Bzdak and Skokov}(2012)}]{Bzdak:2011yy}
\bibinfo{author}{\bibfnamefont{A.}~\bibnamefont{Bzdak}} \bibnamefont{and}
  \bibinfo{author}{\bibfnamefont{V.}~\bibnamefont{Skokov}},
  \bibinfo{journal}{Phys. Lett. B} \textbf{\bibinfo{volume}{710}},
  \bibinfo{pages}{171} (\bibinfo{year}{2012}), \eprint{1111.1949}.

\bibitem[{\citenamefont{Siddique et~al.}(2021)\citenamefont{Siddique, Sheng,
  and Wang}}]{Siddique:2021smf}
\bibinfo{author}{\bibfnamefont{I.}~\bibnamefont{Siddique}},
  \bibinfo{author}{\bibfnamefont{X.-L.} \bibnamefont{Sheng}}, \bibnamefont{and}
  \bibinfo{author}{\bibfnamefont{Q.}~\bibnamefont{Wang}},
  \bibinfo{journal}{Phys. Rev. C} \textbf{\bibinfo{volume}{104}},
  \bibinfo{pages}{034907} (\bibinfo{year}{2021}), \eprint{2106.00478}.

\bibitem[{\citenamefont{Klimenko}(1992)}]{Klimenko:1991he}
\bibinfo{author}{\bibfnamefont{K.~G.} \bibnamefont{Klimenko}},
  \bibinfo{journal}{Z. Phys. C} \textbf{\bibinfo{volume}{54}},
  \bibinfo{pages}{323} (\bibinfo{year}{1992}).

\bibitem[{\citenamefont{Gusynin et~al.}(1994)\citenamefont{Gusynin, Miransky,
  and Shovkovy}}]{Gusynin:1994re}
\bibinfo{author}{\bibfnamefont{V.~P.} \bibnamefont{Gusynin}},
  \bibinfo{author}{\bibfnamefont{V.~A.} \bibnamefont{Miransky}},
  \bibnamefont{and} \bibinfo{author}{\bibfnamefont{I.~A.}
  \bibnamefont{Shovkovy}}, \bibinfo{journal}{Phys. Rev. Lett.}
  \textbf{\bibinfo{volume}{73}}, \bibinfo{pages}{3499} (\bibinfo{year}{1994}),
  \bibinfo{note}{[Erratum: Phys.Rev.Lett. 76, 1005 (1996)]},
  \eprint{hep-ph/9405262}.

\bibitem[{\citenamefont{Gusynin et~al.}(1996)\citenamefont{Gusynin, Miransky,
  and Shovkovy}}]{Gusynin:1995nb}
\bibinfo{author}{\bibfnamefont{V.~P.} \bibnamefont{Gusynin}},
  \bibinfo{author}{\bibfnamefont{V.~A.} \bibnamefont{Miransky}},
  \bibnamefont{and} \bibinfo{author}{\bibfnamefont{I.~A.}
  \bibnamefont{Shovkovy}}, \bibinfo{journal}{Nucl. Phys. B}
  \textbf{\bibinfo{volume}{462}}, \bibinfo{pages}{249} (\bibinfo{year}{1996}),
  \eprint{hep-ph/9509320}.

\bibitem[{\citenamefont{Miransky and Shovkovy}(2002)}]{Miransky:2002rp}
\bibinfo{author}{\bibfnamefont{V.~A.} \bibnamefont{Miransky}} \bibnamefont{and}
  \bibinfo{author}{\bibfnamefont{I.~A.} \bibnamefont{Shovkovy}},
  \bibinfo{journal}{Phys. Rev. D} \textbf{\bibinfo{volume}{66}},
  \bibinfo{pages}{045006} (\bibinfo{year}{2002}), \eprint{hep-ph/0205348}.

\bibitem[{\citenamefont{Endrodi et~al.}(2019)\citenamefont{Endrodi, Giordano,
  Katz, Kov\'acs, and Pittler}}]{Endrodi:2019zrl}
\bibinfo{author}{\bibfnamefont{G.}~\bibnamefont{Endrodi}},
  \bibinfo{author}{\bibfnamefont{M.}~\bibnamefont{Giordano}},
  \bibinfo{author}{\bibfnamefont{S.~D.} \bibnamefont{Katz}},
  \bibinfo{author}{\bibfnamefont{T.~G.} \bibnamefont{Kov\'acs}},
  \bibnamefont{and} \bibinfo{author}{\bibfnamefont{F.}~\bibnamefont{Pittler}},
  \bibinfo{journal}{JHEP} \textbf{\bibinfo{volume}{07}}, \bibinfo{pages}{007}
  (\bibinfo{year}{2019}), \eprint{1904.10296}.

\bibitem[{\citenamefont{Bali et~al.}(2012)\citenamefont{Bali, Bruckmann,
  Endrodi, Fodor, Katz, Krieg, Schafer, and Szabo}}]{Bali:2011qj}
\bibinfo{author}{\bibfnamefont{G.~S.} \bibnamefont{Bali}},
  \bibinfo{author}{\bibfnamefont{F.}~\bibnamefont{Bruckmann}},
  \bibinfo{author}{\bibfnamefont{G.}~\bibnamefont{Endrodi}},
  \bibinfo{author}{\bibfnamefont{Z.}~\bibnamefont{Fodor}},
  \bibinfo{author}{\bibfnamefont{S.~D.} \bibnamefont{Katz}},
  \bibinfo{author}{\bibfnamefont{S.}~\bibnamefont{Krieg}},
  \bibinfo{author}{\bibfnamefont{A.}~\bibnamefont{Schafer}}, \bibnamefont{and}
  \bibinfo{author}{\bibfnamefont{K.~K.} \bibnamefont{Szabo}},
  \bibinfo{journal}{JHEP} \textbf{\bibinfo{volume}{02}}, \bibinfo{pages}{044}
  (\bibinfo{year}{2012}), \eprint{1111.4956}.

\bibitem[{\citenamefont{Fraga and Mizher}(2009)}]{Fraga:2008um}
\bibinfo{author}{\bibfnamefont{E.~S.} \bibnamefont{Fraga}} \bibnamefont{and}
  \bibinfo{author}{\bibfnamefont{A.~J.} \bibnamefont{Mizher}},
  \bibinfo{journal}{Nucl. Phys. A} \textbf{\bibinfo{volume}{820}},
  \bibinfo{pages}{103C} (\bibinfo{year}{2009}), \eprint{0810.3693}.

\bibitem[{\citenamefont{Mizher et~al.}(2010)\citenamefont{Mizher, Chernodub,
  and Fraga}}]{Mizher:2010zb}
\bibinfo{author}{\bibfnamefont{A.~J.} \bibnamefont{Mizher}},
  \bibinfo{author}{\bibfnamefont{M.~N.} \bibnamefont{Chernodub}},
  \bibnamefont{and} \bibinfo{author}{\bibfnamefont{E.~S.} \bibnamefont{Fraga}},
  \bibinfo{journal}{Phys. Rev. D} \textbf{\bibinfo{volume}{82}},
  \bibinfo{pages}{105016} (\bibinfo{year}{2010}), \eprint{1004.2712}.

\bibitem[{\citenamefont{Andersen et~al.}(2016)\citenamefont{Andersen, Naylor,
  and Tranberg}}]{Andersen:2014xxa}
\bibinfo{author}{\bibfnamefont{J.~O.} \bibnamefont{Andersen}},
  \bibinfo{author}{\bibfnamefont{W.~R.} \bibnamefont{Naylor}},
  \bibnamefont{and} \bibinfo{author}{\bibfnamefont{A.}~\bibnamefont{Tranberg}},
  \bibinfo{journal}{Rev. Mod. Phys.} \textbf{\bibinfo{volume}{88}},
  \bibinfo{pages}{025001} (\bibinfo{year}{2016}), \eprint{1411.7176}.

\bibitem[{\citenamefont{Abdallah et~al.}(2022)}]{STAR:2022fan}
\bibinfo{author}{\bibfnamefont{M.}~\bibnamefont{Abdallah}} \bibnamefont{et~al.}
  (\bibinfo{collaboration}{STAR}) (\bibinfo{year}{2022}), \eprint{2204.02302}.

\bibitem[{\citenamefont{Liang and Wang}(2005{\natexlab{b}})}]{Liang:2004xn}
\bibinfo{author}{\bibfnamefont{Z.-T.} \bibnamefont{Liang}} \bibnamefont{and}
  \bibinfo{author}{\bibfnamefont{X.-N.} \bibnamefont{Wang}},
  \bibinfo{journal}{Phys. Lett. B} \textbf{\bibinfo{volume}{629}},
  \bibinfo{pages}{20} (\bibinfo{year}{2005}{\natexlab{b}}),
  \eprint{nucl-th/0411101}.

\bibitem[{\citenamefont{Tang et~al.}(2018)\citenamefont{Tang, Tu, and
  Zhou}}]{Tang:2018qtu}
\bibinfo{author}{\bibfnamefont{A.~H.} \bibnamefont{Tang}},
  \bibinfo{author}{\bibfnamefont{B.}~\bibnamefont{Tu}}, \bibnamefont{and}
  \bibinfo{author}{\bibfnamefont{C.~S.} \bibnamefont{Zhou}},
  \bibinfo{journal}{Phys. Rev. C} \textbf{\bibinfo{volume}{98}},
  \bibinfo{pages}{044907} (\bibinfo{year}{2018}), \eprint{1803.05777}.

\bibitem[{\citenamefont{Yang et~al.}(2018)\citenamefont{Yang, Fang, Wang, and
  Wang}}]{Yang:2017sdk}
\bibinfo{author}{\bibfnamefont{Y.-G.} \bibnamefont{Yang}},
  \bibinfo{author}{\bibfnamefont{R.-H.} \bibnamefont{Fang}},
  \bibinfo{author}{\bibfnamefont{Q.}~\bibnamefont{Wang}}, \bibnamefont{and}
  \bibinfo{author}{\bibfnamefont{X.-N.} \bibnamefont{Wang}},
  \bibinfo{journal}{Phys. Rev. C} \textbf{\bibinfo{volume}{97}},
  \bibinfo{pages}{034917} (\bibinfo{year}{2018}), \eprint{1711.06008}.

\bibitem[{\citenamefont{Xia et~al.}(2021)\citenamefont{Xia, Li, Huang, and
  Zhong~Huang}}]{Xia:2020tyd}
\bibinfo{author}{\bibfnamefont{X.-L.} \bibnamefont{Xia}},
  \bibinfo{author}{\bibfnamefont{H.}~\bibnamefont{Li}},
  \bibinfo{author}{\bibfnamefont{X.-G.} \bibnamefont{Huang}}, \bibnamefont{and}
  \bibinfo{author}{\bibfnamefont{H.}~\bibnamefont{Zhong~Huang}},
  \bibinfo{journal}{Phys. Lett. B} \textbf{\bibinfo{volume}{817}},
  \bibinfo{pages}{136325} (\bibinfo{year}{2021}), \eprint{2010.01474}.

\bibitem[{\citenamefont{Gao}(2021)}]{Gao:2021rom}
\bibinfo{author}{\bibfnamefont{J.-H.} \bibnamefont{Gao}},
  \bibinfo{journal}{Phys. Rev. D} \textbf{\bibinfo{volume}{104}},
  \bibinfo{pages}{076016} (\bibinfo{year}{2021}), \eprint{2105.08293}.

\bibitem[{\citenamefont{M\"uller and Yang}(2022)}]{Muller:2021hpe}
\bibinfo{author}{\bibfnamefont{B.}~\bibnamefont{M\"uller}} \bibnamefont{and}
  \bibinfo{author}{\bibfnamefont{D.-L.} \bibnamefont{Yang}},
  \bibinfo{journal}{Phys. Rev. D} \textbf{\bibinfo{volume}{105}},
  \bibinfo{pages}{L011901} (\bibinfo{year}{2022}), \eprint{2110.15630}.

\bibitem[{\citenamefont{Li and Liu}(2022)}]{Li:2022vmb}
\bibinfo{author}{\bibfnamefont{F.}~\bibnamefont{Li}} \bibnamefont{and}
  \bibinfo{author}{\bibfnamefont{S.~Y.~F.} \bibnamefont{Liu}}
  (\bibinfo{year}{2022}), \eprint{2206.11890}.

\bibitem[{\citenamefont{Wagner et~al.}(2022)\citenamefont{Wagner, Weickgenannt,
  and Speranza}}]{Wagner:2022gza}
\bibinfo{author}{\bibfnamefont{D.}~\bibnamefont{Wagner}},
  \bibinfo{author}{\bibfnamefont{N.}~\bibnamefont{Weickgenannt}},
  \bibnamefont{and} \bibinfo{author}{\bibfnamefont{E.}~\bibnamefont{Speranza}}
  (\bibinfo{year}{2022}), \eprint{2207.01111}.

\bibitem[{\citenamefont{Sheng et~al.}(2020{\natexlab{a}})\citenamefont{Sheng,
  Oliva, and Wang}}]{Sheng:2019kmk}
\bibinfo{author}{\bibfnamefont{X.-L.} \bibnamefont{Sheng}},
  \bibinfo{author}{\bibfnamefont{L.}~\bibnamefont{Oliva}}, \bibnamefont{and}
  \bibinfo{author}{\bibfnamefont{Q.}~\bibnamefont{Wang}},
  \bibinfo{journal}{Phys. Rev. D} \textbf{\bibinfo{volume}{101}},
  \bibinfo{pages}{096005} (\bibinfo{year}{2020}{\natexlab{a}}),
  \bibinfo{note}{[Erratum: Phys.Rev.D 105, 099903 (2022)]},
  \eprint{1910.13684}.

\bibitem[{\citenamefont{Sheng et~al.}(2020{\natexlab{b}})\citenamefont{Sheng,
  Wang, and Wang}}]{Sheng:2020ghv}
\bibinfo{author}{\bibfnamefont{X.-L.} \bibnamefont{Sheng}},
  \bibinfo{author}{\bibfnamefont{Q.}~\bibnamefont{Wang}}, \bibnamefont{and}
  \bibinfo{author}{\bibfnamefont{X.-N.} \bibnamefont{Wang}},
  \bibinfo{journal}{Phys. Rev. D} \textbf{\bibinfo{volume}{102}},
  \bibinfo{pages}{056013} (\bibinfo{year}{2020}{\natexlab{b}}),
  \eprint{2007.05106}.

\bibitem[{\citenamefont{Sheng et~al.}(2022{\natexlab{a}})\citenamefont{Sheng,
  Oliva, Liang, Wang, and Wang}}]{Sheng:2022ffb}
\bibinfo{author}{\bibfnamefont{X.-L.} \bibnamefont{Sheng}},
  \bibinfo{author}{\bibfnamefont{L.}~\bibnamefont{Oliva}},
  \bibinfo{author}{\bibfnamefont{Z.-T.} \bibnamefont{Liang}},
  \bibinfo{author}{\bibfnamefont{Q.}~\bibnamefont{Wang}}, \bibnamefont{and}
  \bibinfo{author}{\bibfnamefont{X.-N.} \bibnamefont{Wang}}
  (\bibinfo{year}{2022}{\natexlab{a}}), \eprint{2206.05868}.

\bibitem[{\citenamefont{Sheng et~al.}(2022{\natexlab{b}})\citenamefont{Sheng,
  Oliva, Liang, Wang, and Wang}}]{Sheng:2022wsy}
\bibinfo{author}{\bibfnamefont{X.-L.} \bibnamefont{Sheng}},
  \bibinfo{author}{\bibfnamefont{L.}~\bibnamefont{Oliva}},
  \bibinfo{author}{\bibfnamefont{Z.-T.} \bibnamefont{Liang}},
  \bibinfo{author}{\bibfnamefont{Q.}~\bibnamefont{Wang}}, \bibnamefont{and}
  \bibinfo{author}{\bibfnamefont{X.-N.} \bibnamefont{Wang}}
  (\bibinfo{year}{2022}{\natexlab{b}}), \eprint{2205.15689}.

\bibitem[{\citenamefont{Nambu and
  Jona-Lasinio}(1961{\natexlab{a}})}]{Nambu:1961fr}
\bibinfo{author}{\bibfnamefont{Y.}~\bibnamefont{Nambu}} \bibnamefont{and}
  \bibinfo{author}{\bibfnamefont{G.}~\bibnamefont{Jona-Lasinio}},
  \bibinfo{journal}{Phys. Rev.} \textbf{\bibinfo{volume}{124}},
  \bibinfo{pages}{246} (\bibinfo{year}{1961}{\natexlab{a}}).

\bibitem[{\citenamefont{Nambu and
  Jona-Lasinio}(1961{\natexlab{b}})}]{Nambu:1961tp}
\bibinfo{author}{\bibfnamefont{Y.}~\bibnamefont{Nambu}} \bibnamefont{and}
  \bibinfo{author}{\bibfnamefont{G.}~\bibnamefont{Jona-Lasinio}},
  \bibinfo{journal}{Phys. Rev.} \textbf{\bibinfo{volume}{122}},
  \bibinfo{pages}{345} (\bibinfo{year}{1961}{\natexlab{b}}).

\bibitem[{\citenamefont{Klimt et~al.}(1990)\citenamefont{Klimt, Lutz, Vogl, and
  Weise}}]{Klimt:1989pm}
\bibinfo{author}{\bibfnamefont{S.}~\bibnamefont{Klimt}},
  \bibinfo{author}{\bibfnamefont{M.~F.~M.} \bibnamefont{Lutz}},
  \bibinfo{author}{\bibfnamefont{U.}~\bibnamefont{Vogl}}, \bibnamefont{and}
  \bibinfo{author}{\bibfnamefont{W.}~\bibnamefont{Weise}},
  \bibinfo{journal}{Nucl. Phys. A} \textbf{\bibinfo{volume}{516}},
  \bibinfo{pages}{429} (\bibinfo{year}{1990}).

\bibitem[{\citenamefont{Vogl et~al.}(1990)\citenamefont{Vogl, Lutz, Klimt, and
  Weise}}]{Vogl:1989ea}
\bibinfo{author}{\bibfnamefont{U.}~\bibnamefont{Vogl}},
  \bibinfo{author}{\bibfnamefont{M.~F.~M.} \bibnamefont{Lutz}},
  \bibinfo{author}{\bibfnamefont{S.}~\bibnamefont{Klimt}}, \bibnamefont{and}
  \bibinfo{author}{\bibfnamefont{W.}~\bibnamefont{Weise}},
  \bibinfo{journal}{Nucl. Phys. A} \textbf{\bibinfo{volume}{516}},
  \bibinfo{pages}{469} (\bibinfo{year}{1990}).

\bibitem[{\citenamefont{Klevansky}(1992)}]{Klevansky:1992qe}
\bibinfo{author}{\bibfnamefont{S.~P.} \bibnamefont{Klevansky}},
  \bibinfo{journal}{Rev. Mod. Phys.} \textbf{\bibinfo{volume}{64}},
  \bibinfo{pages}{649} (\bibinfo{year}{1992}).

\bibitem[{\citenamefont{Buballa}(2005)}]{Buballa:2003qv}
\bibinfo{author}{\bibfnamefont{M.}~\bibnamefont{Buballa}},
  \bibinfo{journal}{Phys. Rept.} \textbf{\bibinfo{volume}{407}},
  \bibinfo{pages}{205} (\bibinfo{year}{2005}), \eprint{hep-ph/0402234}.

\bibitem[{\citenamefont{Volkov and Radzhabov}(2006)}]{Volkov:2005kw}
\bibinfo{author}{\bibfnamefont{M.~K.} \bibnamefont{Volkov}} \bibnamefont{and}
  \bibinfo{author}{\bibfnamefont{A.~E.} \bibnamefont{Radzhabov}},
  \bibinfo{journal}{Phys. Usp.} \textbf{\bibinfo{volume}{49}},
  \bibinfo{pages}{551} (\bibinfo{year}{2006}), \eprint{hep-ph/0508263}.

\bibitem[{\citenamefont{Fukushima}(2008)}]{Fukushima:2008wg}
\bibinfo{author}{\bibfnamefont{K.}~\bibnamefont{Fukushima}},
  \bibinfo{journal}{Phys. Rev. D} \textbf{\bibinfo{volume}{77}},
  \bibinfo{pages}{114028} (\bibinfo{year}{2008}), \bibinfo{note}{[Erratum:
  Phys.Rev.D 78, 039902 (2008)]}, \eprint{0803.3318}.

\bibitem[{\citenamefont{Hatsuda and Kunihiro}(1994)}]{Hatsuda:1994pi}
\bibinfo{author}{\bibfnamefont{T.}~\bibnamefont{Hatsuda}} \bibnamefont{and}
  \bibinfo{author}{\bibfnamefont{T.}~\bibnamefont{Kunihiro}},
  \bibinfo{journal}{Phys. Rept.} \textbf{\bibinfo{volume}{247}},
  \bibinfo{pages}{221} (\bibinfo{year}{1994}), \eprint{hep-ph/9401310}.

\bibitem[{\citenamefont{Liu et~al.}(2015)\citenamefont{Liu, Yu, and
  Huang}}]{Liu:2014uwa}
\bibinfo{author}{\bibfnamefont{H.}~\bibnamefont{Liu}},
  \bibinfo{author}{\bibfnamefont{L.}~\bibnamefont{Yu}}, \bibnamefont{and}
  \bibinfo{author}{\bibfnamefont{M.}~\bibnamefont{Huang}},
  \bibinfo{journal}{Phys. Rev. D} \textbf{\bibinfo{volume}{91}},
  \bibinfo{pages}{014017} (\bibinfo{year}{2015}), \eprint{1408.1318}.

\bibitem[{\citenamefont{Avancini et~al.}(2016)\citenamefont{Avancini, Tavares,
  and Pinto}}]{Avancini:2015ady}
\bibinfo{author}{\bibfnamefont{S.~S.} \bibnamefont{Avancini}},
  \bibinfo{author}{\bibfnamefont{W.~R.} \bibnamefont{Tavares}},
  \bibnamefont{and} \bibinfo{author}{\bibfnamefont{M.~B.} \bibnamefont{Pinto}},
  \bibinfo{journal}{Phys. Rev. D} \textbf{\bibinfo{volume}{93}},
  \bibinfo{pages}{014010} (\bibinfo{year}{2016}), \eprint{1511.06261}.

\bibitem[{\citenamefont{Avancini et~al.}(2017)\citenamefont{Avancini, Farias,
  Benghi~Pinto, Tavares, and Tim\'oteo}}]{Avancini:2016fgq}
\bibinfo{author}{\bibfnamefont{S.~S.} \bibnamefont{Avancini}},
  \bibinfo{author}{\bibfnamefont{R.~L.~S.} \bibnamefont{Farias}},
  \bibinfo{author}{\bibfnamefont{M.}~\bibnamefont{Benghi~Pinto}},
  \bibinfo{author}{\bibfnamefont{W.~R.} \bibnamefont{Tavares}},
  \bibnamefont{and} \bibinfo{author}{\bibfnamefont{V.~S.}
  \bibnamefont{Tim\'oteo}}, \bibinfo{journal}{Phys. Lett. B}
  \textbf{\bibinfo{volume}{767}}, \bibinfo{pages}{247} (\bibinfo{year}{2017}),
  \eprint{1606.05754}.

\bibitem[{\citenamefont{Mao}(2019)}]{Mao:2018dqe}
\bibinfo{author}{\bibfnamefont{S.}~\bibnamefont{Mao}}, \bibinfo{journal}{Phys.
  Rev. D} \textbf{\bibinfo{volume}{99}}, \bibinfo{pages}{056005}
  (\bibinfo{year}{2019}), \eprint{1808.10242}.

\bibitem[{\citenamefont{Coppola et~al.}(2018)\citenamefont{Coppola,
  G\'omez~Dumm, and Scoccola}}]{Coppola:2018vkw}
\bibinfo{author}{\bibfnamefont{M.}~\bibnamefont{Coppola}},
  \bibinfo{author}{\bibfnamefont{D.}~\bibnamefont{G\'omez~Dumm}},
  \bibnamefont{and} \bibinfo{author}{\bibfnamefont{N.~N.}
  \bibnamefont{Scoccola}}, \bibinfo{journal}{Phys. Lett. B}
  \textbf{\bibinfo{volume}{782}}, \bibinfo{pages}{155} (\bibinfo{year}{2018}),
  \eprint{1802.08041}.

\bibitem[{\citenamefont{Chaudhuri et~al.}(2019)\citenamefont{Chaudhuri, Ghosh,
  Sarkar, and Roy}}]{Chaudhuri:2019lbw}
\bibinfo{author}{\bibfnamefont{N.}~\bibnamefont{Chaudhuri}},
  \bibinfo{author}{\bibfnamefont{S.}~\bibnamefont{Ghosh}},
  \bibinfo{author}{\bibfnamefont{S.}~\bibnamefont{Sarkar}}, \bibnamefont{and}
  \bibinfo{author}{\bibfnamefont{P.}~\bibnamefont{Roy}},
  \bibinfo{journal}{Phys. Rev. D} \textbf{\bibinfo{volume}{99}},
  \bibinfo{pages}{116025} (\bibinfo{year}{2019}), \eprint{1907.03990}.

\bibitem[{\citenamefont{Xu et~al.}(2021)\citenamefont{Xu, Chao, and
  Huang}}]{Xu:2020yag}
\bibinfo{author}{\bibfnamefont{K.}~\bibnamefont{Xu}},
  \bibinfo{author}{\bibfnamefont{J.}~\bibnamefont{Chao}}, \bibnamefont{and}
  \bibinfo{author}{\bibfnamefont{M.}~\bibnamefont{Huang}},
  \bibinfo{journal}{Phys. Rev. D} \textbf{\bibinfo{volume}{103}},
  \bibinfo{pages}{076015} (\bibinfo{year}{2021}), \eprint{2007.13122}.

\bibitem[{\citenamefont{Wei et~al.}(2020)\citenamefont{Wei, Jiang, and
  Huang}}]{Wei:2020xfd}
\bibinfo{author}{\bibfnamefont{M.}~\bibnamefont{Wei}},
  \bibinfo{author}{\bibfnamefont{Y.}~\bibnamefont{Jiang}}, \bibnamefont{and}
  \bibinfo{author}{\bibfnamefont{M.}~\bibnamefont{Huang}}
  (\bibinfo{year}{2020}), \eprint{2011.10987}.

\bibitem[{\citenamefont{Yang et~al.}(2022)\citenamefont{Yang, Jin, and
  Hou}}]{Yang:2021hud}
\bibinfo{author}{\bibfnamefont{S.}~\bibnamefont{Yang}},
  \bibinfo{author}{\bibfnamefont{M.}~\bibnamefont{Jin}}, \bibnamefont{and}
  \bibinfo{author}{\bibfnamefont{D.}~\bibnamefont{Hou}},
  \bibinfo{journal}{Chin. Phys. C} \textbf{\bibinfo{volume}{46}},
  \bibinfo{pages}{043107} (\bibinfo{year}{2022}), \eprint{2108.12207}.

\bibitem[{\citenamefont{Miransky and Shovkovy}(2015)}]{Miransky:2015ava}
\bibinfo{author}{\bibfnamefont{V.~A.} \bibnamefont{Miransky}} \bibnamefont{and}
  \bibinfo{author}{\bibfnamefont{I.~A.} \bibnamefont{Shovkovy}},
  \bibinfo{journal}{Phys. Rept.} \textbf{\bibinfo{volume}{576}},
  \bibinfo{pages}{1} (\bibinfo{year}{2015}), \eprint{1503.00732}.

\bibitem[{\citenamefont{Cao}(2021)}]{Cao:2021rwx}
\bibinfo{author}{\bibfnamefont{G.}~\bibnamefont{Cao}}, \bibinfo{journal}{Eur.
  Phys. J. A} \textbf{\bibinfo{volume}{57}}, \bibinfo{pages}{264}
  (\bibinfo{year}{2021}), \eprint{2103.00456}.

\bibitem[{\citenamefont{Brekke and Rosner}(1988)}]{Brekke:1987cc}
\bibinfo{author}{\bibfnamefont{L.}~\bibnamefont{Brekke}} \bibnamefont{and}
  \bibinfo{author}{\bibfnamefont{J.~L.} \bibnamefont{Rosner}},
  \bibinfo{journal}{Comments Nucl. Part. Phys.} \textbf{\bibinfo{volume}{18}},
  \bibinfo{pages}{83} (\bibinfo{year}{1988}).

\bibitem[{\citenamefont{Chang et~al.}(2011)\citenamefont{Chang, Liu, and
  Roberts}}]{Chang:2010hb}
\bibinfo{author}{\bibfnamefont{L.}~\bibnamefont{Chang}},
  \bibinfo{author}{\bibfnamefont{Y.-X.} \bibnamefont{Liu}}, \bibnamefont{and}
  \bibinfo{author}{\bibfnamefont{C.~D.} \bibnamefont{Roberts}},
  \bibinfo{journal}{Phys. Rev. Lett.} \textbf{\bibinfo{volume}{106}},
  \bibinfo{pages}{072001} (\bibinfo{year}{2011}), \eprint{1009.3458}.

\bibitem[{\citenamefont{Fayazbakhsh and Sadooghi}(2014)}]{Fayazbakhsh:2014mca}
\bibinfo{author}{\bibfnamefont{S.}~\bibnamefont{Fayazbakhsh}} \bibnamefont{and}
  \bibinfo{author}{\bibfnamefont{N.}~\bibnamefont{Sadooghi}},
  \bibinfo{journal}{Phys. Rev. D} \textbf{\bibinfo{volume}{90}},
  \bibinfo{pages}{105030} (\bibinfo{year}{2014}), \eprint{1408.5457}.

\bibitem[{\citenamefont{Ayala et~al.}(2016)\citenamefont{Ayala, Dominguez,
  Hernandez, Loewe, and Zamora}}]{Ayala:2015bgv}
\bibinfo{author}{\bibfnamefont{A.}~\bibnamefont{Ayala}},
  \bibinfo{author}{\bibfnamefont{C.~A.} \bibnamefont{Dominguez}},
  \bibinfo{author}{\bibfnamefont{L.~A.} \bibnamefont{Hernandez}},
  \bibinfo{author}{\bibfnamefont{M.}~\bibnamefont{Loewe}}, \bibnamefont{and}
  \bibinfo{author}{\bibfnamefont{R.}~\bibnamefont{Zamora}},
  \bibinfo{journal}{Phys. Lett. B} \textbf{\bibinfo{volume}{759}},
  \bibinfo{pages}{99} (\bibinfo{year}{2016}), \eprint{1510.09134}.

\bibitem[{\citenamefont{'t~Hooft}(1976)}]{tHooft:1976rip}
\bibinfo{author}{\bibfnamefont{G.}~\bibnamefont{'t~Hooft}},
  \bibinfo{journal}{Phys. Rev. Lett.} \textbf{\bibinfo{volume}{37}},
  \bibinfo{pages}{8} (\bibinfo{year}{1976}).

\bibitem[{\citenamefont{Ritus}(1972)}]{Ritus:1972ky}
\bibinfo{author}{\bibfnamefont{V.~I.} \bibnamefont{Ritus}},
  \bibinfo{journal}{Annals Phys.} \textbf{\bibinfo{volume}{69}},
  \bibinfo{pages}{555} (\bibinfo{year}{1972}).

\bibitem[{\citenamefont{Ritus}(1978)}]{Ritus:1978cj}
\bibinfo{author}{\bibfnamefont{V.~I.} \bibnamefont{Ritus}},
  \bibinfo{journal}{Sov. Phys. JETP} \textbf{\bibinfo{volume}{48}},
  \bibinfo{pages}{788} (\bibinfo{year}{1978}).

\bibitem[{\citenamefont{Pauli and Villars}(1949)}]{Pauli:1949zm}
\bibinfo{author}{\bibfnamefont{W.}~\bibnamefont{Pauli}} \bibnamefont{and}
  \bibinfo{author}{\bibfnamefont{F.}~\bibnamefont{Villars}},
  \bibinfo{journal}{Rev. Mod. Phys.} \textbf{\bibinfo{volume}{21}},
  \bibinfo{pages}{434} (\bibinfo{year}{1949}).

\bibitem[{\citenamefont{Carignano and Buballa}(2020)}]{Carignano:2019ivp}
\bibinfo{author}{\bibfnamefont{S.}~\bibnamefont{Carignano}} \bibnamefont{and}
  \bibinfo{author}{\bibfnamefont{M.}~\bibnamefont{Buballa}},
  \bibinfo{journal}{Phys. Rev. D} \textbf{\bibinfo{volume}{101}},
  \bibinfo{pages}{014026} (\bibinfo{year}{2020}), \eprint{1910.03604}.

\bibitem[{\citenamefont{Kapusta and Gale}(2011)}]{Kapusta:2006pm}
\bibinfo{author}{\bibfnamefont{J.~I.} \bibnamefont{Kapusta}} \bibnamefont{and}
  \bibinfo{author}{\bibfnamefont{C.}~\bibnamefont{Gale}},
  \emph{\bibinfo{title}{{Finite-temperature field theory: Principles and
  applications}}}, Cambridge Monographs on Mathematical Physics
  (\bibinfo{publisher}{Cambridge University Press}, \bibinfo{year}{2011}), ISBN
  \bibinfo{isbn}{978-0-521-17322-3, 978-0-521-82082-0, 978-0-511-22280-1}.

\bibitem[{\citenamefont{MOTT}(1968)}]{Mott:1968nwb}
\bibinfo{author}{\bibfnamefont{N.~F.} \bibnamefont{MOTT}},
  \bibinfo{journal}{Rev. Mod. Phys.} \textbf{\bibinfo{volume}{40}},
  \bibinfo{pages}{677} (\bibinfo{year}{1968}).

\bibitem[{\citenamefont{Hufner et~al.}(1996)\citenamefont{Hufner, Klevansky,
  and Rehberg}}]{Hufner:1996pq}
\bibinfo{author}{\bibfnamefont{J.}~\bibnamefont{Hufner}},
  \bibinfo{author}{\bibfnamefont{S.~P.} \bibnamefont{Klevansky}},
  \bibnamefont{and} \bibinfo{author}{\bibfnamefont{P.}~\bibnamefont{Rehberg}},
  \bibinfo{journal}{Nucl. Phys. A} \textbf{\bibinfo{volume}{606}},
  \bibinfo{pages}{260} (\bibinfo{year}{1996}).

\bibitem[{\citenamefont{Costa et~al.}(2003)\citenamefont{Costa, Ruivo, and
  Kalinovsky}}]{Costa:2002gk}
\bibinfo{author}{\bibfnamefont{P.}~\bibnamefont{Costa}},
  \bibinfo{author}{\bibfnamefont{M.~C.} \bibnamefont{Ruivo}}, \bibnamefont{and}
  \bibinfo{author}{\bibfnamefont{Y.~L.} \bibnamefont{Kalinovsky}},
  \bibinfo{journal}{Phys. Lett. B} \textbf{\bibinfo{volume}{560}},
  \bibinfo{pages}{171} (\bibinfo{year}{2003}), \eprint{hep-ph/0211203}.

\bibitem[{\citenamefont{Blaschke et~al.}(2017)\citenamefont{Blaschke, Dubinin,
  Radzhabov, and Wergieluk}}]{Blaschke:2016sqn}
\bibinfo{author}{\bibfnamefont{D.}~\bibnamefont{Blaschke}},
  \bibinfo{author}{\bibfnamefont{A.}~\bibnamefont{Dubinin}},
  \bibinfo{author}{\bibfnamefont{A.}~\bibnamefont{Radzhabov}},
  \bibnamefont{and}
  \bibinfo{author}{\bibfnamefont{A.}~\bibnamefont{Wergieluk}},
  \bibinfo{journal}{Phys. Rev. D} \textbf{\bibinfo{volume}{96}},
  \bibinfo{pages}{094008} (\bibinfo{year}{2017}), \eprint{1608.05383}.

\bibitem[{\citenamefont{Mao}(2021)}]{Mao:2019avr}
\bibinfo{author}{\bibfnamefont{S.}~\bibnamefont{Mao}}, \bibinfo{journal}{Chin.
  Phys. C} \textbf{\bibinfo{volume}{45}}, \bibinfo{pages}{021004}
  (\bibinfo{year}{2021}), \eprint{1908.02851}.

\bibitem[{\citenamefont{Ishikawa et~al.}(2005)}]{Ishikawa:2004id}
\bibinfo{author}{\bibfnamefont{T.}~\bibnamefont{Ishikawa}}
  \bibnamefont{et~al.}, \bibinfo{journal}{Phys. Lett. B}
  \textbf{\bibinfo{volume}{608}}, \bibinfo{pages}{215} (\bibinfo{year}{2005}),
  \eprint{nucl-ex/0411016}.

\bibitem[{\citenamefont{Muto et~al.}(2007)}]{KEK-PS-E325:2005wbm}
\bibinfo{author}{\bibfnamefont{R.}~\bibnamefont{Muto}} \bibnamefont{et~al.}
  (\bibinfo{collaboration}{KEK-PS-E325}), \bibinfo{journal}{Phys. Rev. Lett.}
  \textbf{\bibinfo{volume}{98}}, \bibinfo{pages}{042501}
  (\bibinfo{year}{2007}), \eprint{nucl-ex/0511019}.

\bibitem[{\citenamefont{Qian et~al.}(2009)}]{CLAS:2009kjz}
\bibinfo{author}{\bibfnamefont{X.}~\bibnamefont{Qian}} \bibnamefont{et~al.}
  (\bibinfo{collaboration}{CLAS}), \bibinfo{journal}{Phys. Lett. B}
  \textbf{\bibinfo{volume}{680}}, \bibinfo{pages}{417} (\bibinfo{year}{2009}),
  \eprint{0907.2668}.

\bibitem[{\citenamefont{Foldy and Wouthuysen}(1950)}]{Foldy:1949wa}
\bibinfo{author}{\bibfnamefont{L.~L.} \bibnamefont{Foldy}} \bibnamefont{and}
  \bibinfo{author}{\bibfnamefont{S.~A.} \bibnamefont{Wouthuysen}},
  \bibinfo{journal}{Phys. Rev.} \textbf{\bibinfo{volume}{78}},
  \bibinfo{pages}{29} (\bibinfo{year}{1950}).

\bibitem[{\citenamefont{Dothan}(1982)}]{Dothan:1981ex}
\bibinfo{author}{\bibfnamefont{Y.}~\bibnamefont{Dothan}},
  \bibinfo{journal}{Physica A} \textbf{\bibinfo{volume}{114}},
  \bibinfo{pages}{216} (\bibinfo{year}{1982}).

\bibitem[{\citenamefont{Kawaguchi and Huang}(2022)}]{Kawaguchi:2022dbq}
\bibinfo{author}{\bibfnamefont{M.}~\bibnamefont{Kawaguchi}} \bibnamefont{and}
  \bibinfo{author}{\bibfnamefont{M.}~\bibnamefont{Huang}}
  (\bibinfo{year}{2022}), \eprint{2205.08169}.

\end{thebibliography}

\end{document}